\documentclass[longauth]{aa}  
\usepackage{epsfig}
\usepackage{amsmath}
\usepackage{amssymb}
\usepackage{natbib}
\usepackage{hyperref}
\usepackage{graphicx}
\usepackage{times}
  \usepackage[flushleft]{threeparttable}
  
\usepackage{CJKutf8}
\usepackage{xcolor}%\usepackage{helvetica} % uses helvetica postscript font (download helvetica.sty)

\bibpunct{(}{)}{;}{a}{}{,}             
\newcommand{\python}{{\tt Python}}

\defcitealias{KM05}{KM05}
\defcitealias{PN11}{PN11}
\defcitealias{HC11}{HC11}
\defcitealias{FK12}{FK12}
\defcitealias{burkmocz}{BM19}

\begin{document}

\title{Reconciling extragalactic star formation efficiencies with theory: insights from PHANGS}
%\title{Matching turbulence-regulated star formation models to extragalactic cloud populations: insights from PHANGS}

\author{
Sharon~E.~Meidt\inst{1}   \and
Simon~C.~O.~Glover\inst{2}   \and
Ralf~S.~Klessen\inst{2,3,4,5}   \and
Adam~K.~Leroy\inst{6}   \and
%Jiayi~Sun (孙嘉懿) \inst{7,8}   \and%\begin{CJK*}{UTF8}{gbsn}(孙嘉懿)\end{CJK*}\inst{7,8}   \and
Jiayi~Sun\inst{7,8}   \and%\begin{CJK*}{UTF8}{gbsn}(孙嘉懿)\end{CJK*}\inst{7,8}   \and
Oscar Agertz\inst{9}   \and
Eric Emsellem\inst{10,11}    \and
Jonathan~D.~Henshaw\inst{12}    \and
Lukas Neumann\inst{10,13}    \and
Erik~Rosolowsky\inst{14}    \and
Eva~Schinnerer\inst{15}   \and
Dyas Utomo\inst{6}    \and
Arjen van der Wel\inst{1}    \and
Frank Bigiel \inst{16}    \and
Dario Colombo\inst{16}    \and
Damian R. Gleis\inst{15}    \and
Kathryn~Grasha\inst{17,18,19}    \and
Jindra Gensior\inst{20}    \and
Oleg Y. Gnedin\inst{21}    \and
Annie~Hughes\inst{22,23}    \and
Eric J. Murphy\inst{24}    \and
Miguel~Querejeta\inst{25}    \and
Rowan J. Smith\inst{26}    \and
Thomas G. Williams\inst{27}    \and
Antonio~Usero\inst{25} 
}

\institute{
%1
Sterrenkundig Observatorium, Universiteit Gent, Krijgslaan 281 S9, B-9000 Gent, Belgium  \and
%2
Universit\"{a}t Heidelberg, Zentrum f\"{u}r Astronomie, Institut f\"{u}r Theoretische Astrophysik, Albert-Ueberle-Str 2, D-69120 Heidelberg, Germany   \and
%3
Universit\"{a}t Heidelberg, Interdisziplin\"{a}res Zentrum f\"{u}r Wissenschaftliches Rechnen, Im Neuenheimer Feld 225, 69120 Heidelberg, Germany  \and
%4
Harvard-Smithsonian Center for Astrophysics, 60 Garden Street, Cambridge, MA 02138, U.S.A.  \and
%5
Elizabeth S. and Richard M. Cashin Fellow at the Radcliffe Institute for Advanced Studies at Harvard University, 10 Garden Street, Cambridge, MA 02138, U.S.A.  \and
%6
Department of Astronomy, The Ohio State University, 140 West 18th Avenue, Columbus, OH 43210, USA  \and
%7
NASA Hubble Fellow  \and
%8
Department of Astrophysical Sciences, Princeton University, 4 Ivy Lane, Princeton, NJ 08544, USA  \and
%9
Lund Observatory, Division of Astrophysics, Department of Physics, Lund University, Box 43, SE-221 00 Lund, Sweden  \and
%10
European Southern Observatory, Karl-Schwarzschild Stra{\ss}e 2, D-85748 Garching bei M\"{u}nchen, Germany  \and
%11
Univ Lyon, Univ Lyon1, Ens de Lyon, CNRS, Centre de Recherche Astrophysique de Lyon UMR5574, F-69230, Saint-Genis-Laval, France  \and
%12
Astrophysics Research Institute, Liverpool John Moores University, 146 Brownlow Hill, Liverpool L3 5RF, UK  \and
%13
Argelander-Institut f\"ur Astronomie, Universit\"at Bonn, Auf dem H\"ugel 71, 53121 Bonn, Germany  \and
%14
Department of Physics, University of Alberta, Edmonton, AB T6G 2E1, Canada  \and
%15
Max-Planck-Institut f\"{u}r Astronomie, K\"{o}nigstuhl 17, D-69117, Heidelberg, Germany  \and
%16
Argelander-Institut f\"ur Astronomie, Universit\"at Bonn, Auf dem H\"ugel 71, 53121 Bonn, Germany  \and
%17
ARC DECRA Fellow  \and
%18
Research School of Astronomy and Astrophysics, Australian National University, Canberra, ACT 2611, Australia  \and
%19
ARC Centre of Excellence for All Sky Astrophysics in 3 Dimensions (ASTRO 3D), Australia  \and
%20
Institute for Astronomy, University of Edinburgh, Royal Observatory, Blackford Hill, Edinburgh EH9 3HJ, UK  \and
%21
Department of Astronomy, University of Michigan, Ann Arbor, MI 48109, USA  \and
%22
CNRS, IRAP, 9 Av. du Colonel Roche, BP 44346, F-31028 Toulouse cedex 4, France  \and
%23
Universit\'{e} de Toulouse, UPS-OMP, IRAP, F-31028 Toulouse cedex 4, France  \and
%24
NRAO National Radio Astronomy Observatory, 520 Edgemont Road, Charlottesville, VA 22903  \and
%25
Observatorio Astron\'omico Nacional (IGN), C/ Alfonso XII, 3, E-28014 Madrid, Spain  \and
%26
SUPA, School of Physics and Astronomy, University of St Andrews, North Haugh, St Andrews, KY16 9SS  \and
%27
Sub-department of Astrophysics, Department of Physics, University of Oxford, Keble Road, Oxford OX1 3RH, UK
}
 \date{December 20, 2024}

\abstract
{New extragalactic measurements of the cloud population-averaged star formation efficiency per free-fall time $\rm \epsilon_{\rm ff}$ from PHANGS show little sign of a theoretically predicted dependence on the gas virial level and weak variation with cloud-scale gas velocity dispersion.  
We explore ways to bring theory into consistency with the observations, in particular highlighting systematic variations in internal density structure that must happen together with an increase in virial parameter typical towards denser galaxy centers.  
To introduce these variations into conventional turbulence-regulated star formation models, we adopt three adjustments, together motivated by the expectation that the background host galaxy has an influence on the cloud-scale: we incorporate self-gravity and an internal density distribution that contains a broad power-law (PL) component and resembles the structure observed in local resolved clouds, we allow that the internal gas kinematics include motion in the background potential and let this regulate the onset of self-gravitation, and we assume that the distribution of gas densities is in a steady-state for only a fraction of a cloud free-fall time.  In practice, these changes significantly reduce the efficiencies predicted in multi-free-fall (MFF) scenarios compared to purely lognormal (LN) probability density functions (PDFs) and ties efficiency variations to variations in the slope of the PL $\alpha$.  
We fit the model to PHANGS measurements of $\epsilon_{\rm ff}$ to identify the PL slopes that yield an optimal match.  These slopes vary systematically with galactic environment in the sense that gas that sits furthest from virial balance contains fractionally more gas at high density.  We relate this to the equilibrium response of gas in the presence of the galactic gravitational potential, which forces more gas to high density than characteristic of fully self-gravitating clouds.   
Viewing the efficiency variations as originating with time evolution in the power-law slope, our findings would alternatively imply coordination of the cloud evolutionary stage within environment.   
With this `galaxy regulation' behavior included, our preferred `self-gravitating' sgMFF models function similarly to the original, roughly `virialized cloud' single-free-fall models.  However, outside the environment of disks with their characteristic regulation, the flexible MFF models may be better suited.   
}
   \keywords{galaxies:ISM, galaxies: star formation}

    \titlerunning{
    Turbulence regulated star formation: insights from PHANGS}
    \authorrunning{Meidt et al.}
   \maketitle

\section{Introduction}
\label{sec:intro}
\setcounter{footnote}{0}
Our view of the way stars form out of cold dense gas has grown ever more precise in recent years, with observations and numerical simulations capturing an expanding range of the spatial scales involved in the process.  We now see hints that the star-forming activity inside individual star-forming cloud structures is related to conditions on the cloud scale, and that these `initial conditions' for star formation are inherited from the conditions and processes present on even larger scales in the host galaxy.  The molecular cloud interface between the small and large (host galaxy) scales, in particular, is an indispensable source of information on how large-scale mechanisms guide gas organization 
and regulate cloud formation and destruction processes that influence the properties of star-forming clouds \citep[e.g.][]{hughes13,DobbsPringle13, colombo14, meidt15, Duarte-Cabral16,Dobbs19,Chevance20,tress20,smith20,pettitt20,henshaw20,Querejeta21,barnes21,meidt21,choi23,choi24,lu24,schinnerLeroyReview}.  

One of the key constraints in this modern view is the non-thermal (turbulent) motion present on the cloud scale, which allows us to anchor the turbulence regulation at the heart of our theory for star formation \citep{KM05,PN11,HC11,FK12} aimed toward reproducing the local rates at which gas forms stars in any context, across cosmic time.  In this model, the turbulent properties of the gas driven on the cloud-scale determine both the density structure enclosed within self-gravitating clouds, and the internal competition provided against gravity that regulates the formation of star-forming cores \citep{maclow04,mckeeOstriker07,klessenglover,girichidis20}.  

Although these models can reproduce to zeroth order the inefficiency of star formation \citep[as pointed out e.g.\ by][]{zuckermanEvans74}, they are in tension with our best constraints on the star forming efficiencies of extragalactic cloud populations, which exhibit trends with turbulent gas motions and level of virialization not fully captured by the models \citep{leroy17, utomo18, sun23, schinnerLeroyReview, leroy24}.

Fortunately, the richness of data sets like the Physics at High Angular resolution in Nearby GalaxieS\footnote{\url{https://sites.google.com/view/phangs/home}} (PHANGS) survey, which samples the star formation cycle in a diversity of environments, provide insight into how turbulence-regulated star formation models can be tailored to match observations. For example, extragalactic cloud-scale observations consistently show that many clouds are in a slightly super-virial state \citep{sun18,sun20a, rosolowsky21,evans22} that in turn appears sensitive to the host galaxy environment \citep[together with other properties;][]{hughes13, colombo14,rosolowsky21}.  
In galaxy centers, enhanced stellar gravity \citep[e.g.][]{meidt18}, strong shear \citep[e.g][]{liu21,lu24} and typically complex kinematics \citep{henshaw20,choi23,henshawinprep} are thought to contribute to the observed excess of kinetic energy.   Galaxy centers are also locations with strong radiation fields and magnetic field strengths.  All these factors can in turn make a large portion of the cloud inert and determine where in the cloud interior the gas can become self-gravitating and undergo gravitational collapse \citep{meidt20}.  

The creation of an inert molecular component is also predicted to be a consequence of feedback from star formation.  SNe are capable of driving turbulence that regulates the dynamical state of the gas across a range of scales \citep{padoan16, ostriker10}, while earlier feedback in the form of stellar winds, radiation pressure and photoionization leads to rapid cloud destruction \citep{kim18,Kruijssen19, Chevance20} that both cuts off the star formation happening within individual clouds and drives a cycle in which molecular gas is forced into periods of quiescence \citep{semenov17,semenov18}.  

While increases in virial level are generally expected to reduce the measured star formation efficiency \citep[e.g.][]{Dobbs11, padoan14,padoan16}, modern extragalactic observational censuses less obviously show such a link \citep{leroy24}.  
Our goal in this paper is to identify factors that can improve the match between turbulence-regulated star formation theories and the behavior observed in galactic and extragalactic cloud populations. 
We consider a picture in which the internal structure and star-forming ability of clouds are regulated to some extent by conditions outside of them; as the gas in increasingly pressurized (central) environments becomes super-virial, it must enclose more and more dense gas, thereby preventing the efficiency from decreasing substantially.  

We introduce this behavior in practice by starting with the principles of turbulence-regulated core formation in the \citet[][hereafter KM05]{KM05}, \citet[][hereafter PN11]{PN11} and \citet[][hereafter HC11]{HC11} models but replacing a pure lognormal (LN) density PDF with a hybrid one that includes a power-law (PL) tail to explicitly account for gas self-gravity. 
Thus our fiducial model behaves similarly to the model proposed by \cite{burk18} and \citet[hereafter BM19]{burkmocz}, in which variations in efficiency are primarily the result of variations in the slope of the PL tail.  Here, though, the density PDF is revised so that the PL tail can be as shallow as observed in local clouds \citep{alves17,Kainulainen14,schneider22} but start at the threshold where gas self-gravity dominates over the external galactic potential as proposed by \cite{meidt20}.  We also restrict core formation to the material that is self-gravitating, rather than allowing all cloud material (including the LN component) to participate in a multi-free-fall process. 
As a result of these two modifications, so-called `multi-free-fall' scenarios \citep[][hereafter FK12]{FK12} are able to yield low star formation efficiencies as a consequence of low-efficiency core formation, as first postulated by \citetalias{KM05}, without an ad hoc lowering of the core-to-star efficiency.  

After describing new extragalactic measurements of the efficiency of star formation per free-fall time $\rm \epsilon_{\rm ff}$ (sometimes called $\rm SFE_{\rm ff}$) made by \cite{leroy24} using data from the PHANGS survey in \S\ref{sec:PHANGS_SFES}, we summarize the features of the main classes of turbulence-regulated SF models in \S\ref{sec:theorycomp}.  We then describe our proposed modification to these models in \S\ref{sec:proposedmodifications} and relate this to the influence of galactic environment in \S\ref{sec:M20bottleneck}.  Then in \S\ref{sec:obsTests} we compare our modified model of turbulence-regulated core formation to $\epsilon_{\rm ff}$ measured by PHANGS, using the approach summarized in \S\ref{sec:strategy} and described in more detail in \S\ref{sec:empiricalTests}, based on constraints from PHANGS on the cloud-scale properties of gas and the local galactic environment summarized in \S\ref{sec:PHANGS}.  

\section{Variations in extragalactic star formation efficiencies}\label{sec:PHANGS_SFES}%: variation with environment-sensitive cloud-scale gas properties}
\subsection{Overview}
The observational access to molecular gas properties on the cloud scale in nearby galaxies \citep{schinnerer13, leroy17, leroy21,schinnerLeroyReview} has marked a shift in our empirical description of the star formation process.  As demonstrated by \cite{leroy17}, \cite{utomo18} and \cite{leroy24}, we can now constrain the star formation efficiency per free-fall time $\epsilon_{\rm ff}$ in extragalactic targets (see definition in Eq. [\ref{eq:obsSFE}]), making it possible to compare directly with the predictions of theory and with galactic observations across a wide range of complementary environmental conditions.

The early studies leveraging this technique found signs of local and global variation in $\rm \epsilon_{\rm ff}$, both as a function of cloud-scale gas properties and from galaxy to galaxy \citep[e.g.][]{utomo18,Schruba19}.  Since those studies, PHANGS has widened the environments and targets where we have $\rm \epsilon_{\rm ff}$ measurements and incorporated modern empirical constraints on the CO-to-H$_2$ conversion factor \citep[see][for a review]{schinnerLeroyReview}.  
As closely examined by \cite{leroy24}, we now have a clearer picture of how local environmental conditions impact cloud-scale gas properties \citep[see also][]{rosolowsky21, sun20a} and influence the rate of star formation within clouds.  Taking measurements from 67 galaxies in total sampling 841 regions with high CO completeness, \cite{leroy24} find a strong correlation between the molecular gas depletion time measured on kpc-scales and the cloud (150-pc) scale density, which in turn implies little systematic variation in $\rm \epsilon_{\rm ff}$ (see Eq. \ref{eq:obsSFE}).  
Both the conversion factor and the completeness correction adopted by \cite{leroy24} are key to recovering this behavior. With these choices, the $\rm \epsilon_{\rm ff}$ exhibits little sign of the anticorrelation with cloud velocity dispersion or virial parameter hinted at in earlier empirical studies, which would put extragalactic observations into strong conflict with turbulence-regulated star formation theories \citep{sun23, schinnerLeroyReview}.  However, not all tension with the models is alleviated; the theoretical dependence of $\rm \epsilon_{\rm ff}$ on cloud properties, and cloud dynamical state in particular, fails to provide a satisfactory match to the PHANGS measurements.  

In this paper, we examine the sensitivity of $\rm \epsilon_{\rm ff}$ to cloud-scale boundary conditions in turbulence-regulated star formation models and in particular consider the extent to which environment impacts these conditions.  
Our main point of reference is the set of star formation models proposed by \citetalias{KM05}, \citetalias{PN11}, \citetalias{HC11}, \citetalias{FK12} \& \citetalias{burkmocz}, which we discuss in \S\ref{sec:mainfeatures}.  
Before comparing to those models, we first describe the methodology used to measure $\rm \epsilon_{\rm ff}$ in our PHANGS targets.  Later we use the comparison to motivate modifications to turbulence-regulated star formation theories that can help improve the quality of the agreement between the models and the observations.  
\subsection{Reconstruction of the cloud-scale star formation efficiency per free-fall time $\rm \epsilon_{\rm ff}$}\label{sec:phangssfes}
\subsubsection{A cloud population-averaged (time-averaged) view of the star formation cycle}%\label{sec:phangsmeasurements}
The information needed to test the cloud-scale factors influencing the process of star formation is currently accessible in the PHANGS high-level measurement database presented by \citet[updated by \citealt{sun23}]{sun22} and used by \cite{leroy24}.    In this paper, we employ a number of multi-wavelength measurements from throughout the database, including star formation rates (SFRs), gas surface densities and the free-fall times in some 841 regions sampling throughout the disks of 67 PHANGS galaxies.  We refer the reader to \cite{sun23} and \cite{leroy24} for details.  

Briefly, measurements are extracted in hexagonal regions 
1.5~kpc in size and encompass many individual molecular clouds participating in the star formation process at different moments in the star formation cycle \citep[i.e.][]{Kruijssen14,semenov18}.  Following \cite{leroy24}, as well as \cite{leroy17}, \cite{utomo18} and \cite{sun22}, each kpc-scale SFR measurement is matched with a kpc-scale molecular gas surface density $\Sigma_{\rm mol}^{\rm kpc}$.  
These kpc-scale measurements are then matched with measurements of the typical cloud-scale gas properties within each kpc-size region (see more below).  The latter are essential for constraining the gas free-fall time, which makes it possible with this approach to observationally reconstruct the efficiency per free-fall time as 
\begin{equation}
\rm \epsilon_{\rm ff}^{\rm obs}=t_{\rm ff} \frac{\Sigma_{\rm SFR}^{\rm kpc}}{\Sigma_{\rm mol}^{\rm kpc}}\label{eq:obsSFE}
\end{equation}
where $\Sigma_{\rm mol}^{\rm kpc}$ and $\Sigma_{\rm SFR}^{\rm kpc}$ are the mean molecular gas and star formation rate surface densities over the whole  %$0.5{-}
$1.5$-kpc averaging aperture\footnote{Following \cite{sun23} and \citet{leroy24}, these measurements are extracted from maps of $\Sigma_{\rm mol}$ and $\Sigma_{\rm SFR}$ that have been convolved to share a matched Gaussian kernel with FWHM equal to the hex diameter and then sampled at the hex center.} 
and the free-fall time 
\begin{equation}
t_{\rm ff}=\left(\frac{3\pi}{32G\rho}\right)^{1/2}\label{eq:tff}
\end{equation} 
is estimated using the observationally reconstructed mass-weighted average cloud-scale gas volume density $\langle\rho_{\rm mol}^{\rm cloud}\rangle$ in the kpc-size region estimated from the cloud-scale gas surface density $\langle\Sigma_{\rm mol}^{\rm cloud}\rangle$ and vertical size $h$.  Following \cite{sun23,leroy24} we assume a constant vertical size $h=100$~pc. 

For estimating both $\langle\Sigma_{\rm mol}^{\rm cloud}\rangle$ and $\Sigma_{\rm mol}^{\rm kpc}$ in this work, we follow \cite{leroy24} and adopt the new CO-to-H$_2$ conversion factor $\alpha_{\rm CO}^{2-1}$ recommended by \citet[][]{schinnerLeroyReview}.  We also take into account the completeness correction advised by \citet{leroy24} and developed by \cite{sun23}, which retains only regions where a high fraction of the total CO flux in each 1.5~kpc region, $f_{\rm comp}$, is recovered in high resolution interferometric observations, and hence is reflected in measurements of cloud-scale gas properties based on these observations.  
Following \cite{leroy24} we select these `high completeness' regions with the threshold criteria $f_{\rm comp}>0.5$ and $\langle\Sigma_{\rm mol}^{\rm cloud}\rangle>20$~$M_\odot$~pc$^{-2}$.  This yields $841$ regions sampling across the targets set of $67$ galaxies.  The reader is referred to \cite{leroy24} for a detailed discussion of how the conversion factor and completeness corrections impact the measured efficiencies.  Those choices will not be discussed here.  

It should be noted \citep[and see also][]{leroy17,sun22} that, given the clumpiness of molecular gas, the kpc-averaged molecular gas surface density in a given region $\Sigma_{\rm gas}^{\rm kpc}$ used in Eq.~(\ref{eq:obsSFE}) is not the same as the mean of the molecular gas surface densities of the clouds within that 1.5-kpc region $\langle\Sigma_{\rm mol}^{\rm cloud}\rangle$.  
Reconstructions of $\epsilon_{\rm ff}$ based on $\langle\Sigma_{\rm mol}^{\rm cloud}\rangle$ (instead of $\Sigma_{\rm mol}^{\rm kpc}$) would require measurements of the SFR  at an equivalently high cloud-scale resolution \citep[i.e.~to capture the expected clumpiness of the distribution of recent star formation;][]{sun22}.  
Such SFR maps would not necessarily furnish more realistic $\epsilon_{\rm ff}$ estimates representative of the local star-forming cycle, though, given the timescales associated with most extragalactic star formation tracers \citep[see also][]{grudic19}.  With the exception of an embedded phase, most probes of recent star formation are spatially decoupled from the gas, as a result of the nature of the star formation cycle \citep{Kruijssen14,Chevance20}. 
For our purposes, $\rm \epsilon_{\rm ff}^{\rm obs}$ measured as in Eq. (\ref{eq:obsSFE}) on kpc scales is thus preferred.  For a population of roughly identical clouds observed in a given kpc-scale region, $\rm \epsilon_{\rm ff}$ in Eq. (\ref{eq:obsSFE}) is a good approximation of the time-average $\epsilon_{\rm ff}$ that occurs throughout the individual clouds, provided that the sub-1.5 kpc distribution of star formation is of comparable clumpiness to that of the gas \citep[a factor of roughly 2;][]{sun22}.  

\subsubsection{Cloud-scale gas properties}\label{sec:gasprops}
Each 1.5 kpc sized hexagonal zone in the measurement grid for a given galaxy is assigned a `representative cloud-scale pixel' whose properties are defined by the mass-weighted average properties of the cloud-scale regions probed in our PHANGS-ALMA CO(2-1) maps within that zone \citep[see][for more details]{sun18,sun20a,leroy24}.  The gas properties are measured on a fixed physical scale, and we adopt a beam full-width-half-maximum scale corresponding to $D_{beam}=150$~pc as our fiducial measurement scale.  
In this case, we assign a 2D cloud radius $R_c=150/2$~pc for the radius of the representative cloud in the plane of the disk and follow \cite{sun23} and \cite{leroy24} and adopt a fixed molecular gas scale height $h = 100$~pc. We also follow \cite{sun22} and assign a three-dimensional mean radius for each cloud scale pixel measured in terms of the beam size and our adopted scale height $h = 100$~pc.  For each representative cloud-scale pixel, measured properties include the mass-weighted average surface density $\langle\Sigma_{\rm mol}^{\rm cloud}\rangle$, mass $\langle M_{\rm mol}^{\rm cloud}\rangle$, virial parameter $\langle \alpha_{\rm vir}^{\rm cloud}\rangle$ and velocity dispersion $\langle\sigma_{\rm mol}^{\rm cloud}\rangle$, which we use to estimate the turbulent Mach number assuming a sound speed $c_s = 0.3$~km~s$^{-1}$ (corresponding to a temperature of 20 K). Note that although we might expect the cloud-averaged gas temperature to vary significantly as a function of the strength of the local interstellar radiation field, the temperature of the highly-shielded molecular phase traced by CO shows much less sensitivity to the radiation field strength \citep[see e.g.][]{penaloza}, justifying our use of a constant value here. 

With these observables we determine a number of additional properties for each `representative cloud-scale pixel', such as the gas virial state and the strength of gas self-gravity on the cloud scale (the cloud potential) for comparison to the galactic potential estimated as described in $\S$ \ref{sec:decouplethresh}. %Appendix \ref{sec:appendixgamma} (using a scaling factor calculated in Appendix \ref{sec:appendixbk}).  
Again, we use the fixed molecular gas scale height $h = 100$~pc to determine the gas volume density $\langle\rho_{\rm mol}^{\rm cloud}\rangle$=$\langle\Sigma_{\rm mol}^{\rm cloud}\rangle/2h$.  We then substitute this value into the denominator of Eq. (\ref{eq:tff}) to estimate the cloud free-fall time needed to calculate the efficiency per free-fall time from the observations.

We refer the reader to \cite{leroy24} for a discussion of the considerations (including the CO-to-H$_2$ conversion factor) that influence the measured $t_{\rm ff}$ and $\epsilon_{\rm ff}$, and to \cite{sun22} for the impact of the vertical size and the choice of clouds vs.\ pixels.  We also refer the reader to \cite{sun20a} for an assessment of cloud-scale gas properties compared to those in the PHANGS GMC catalogs.

\begin{figure*}[t]
%\begin{flushleft}
\vspace*{-.5in}
\begin{center}
\begin{tabular}{c}
\includegraphics[width=0.95\linewidth]{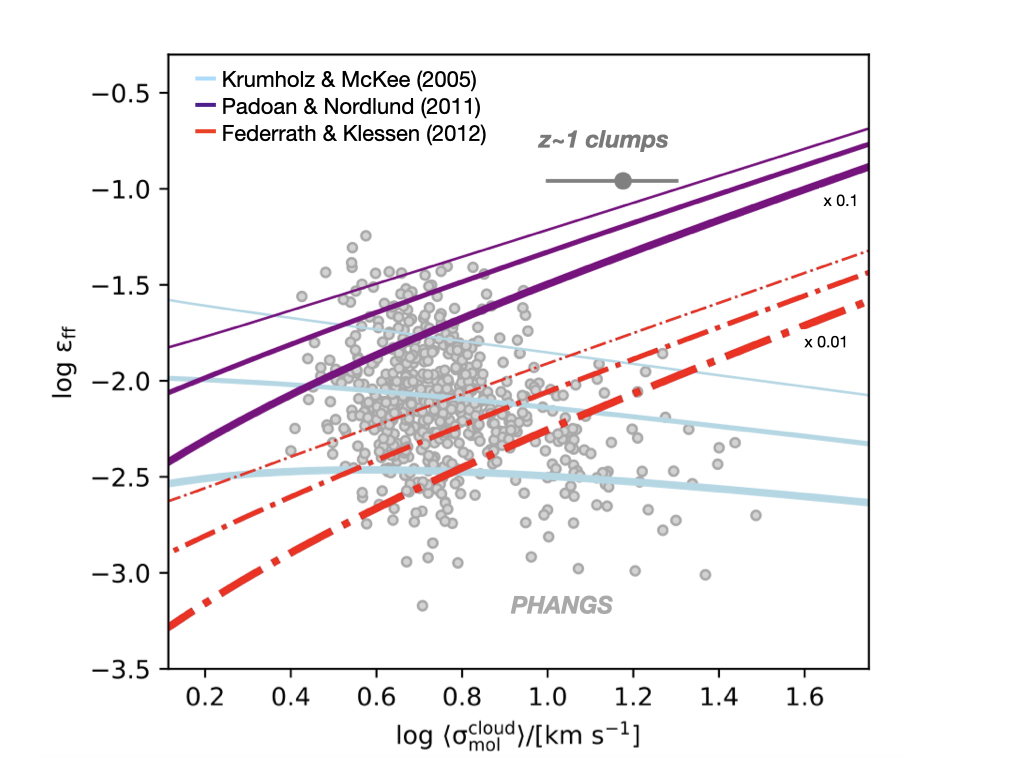}%phangs_models.jpg}%modBplot150nh.jpg}%pdf}
%\vspace*{-.5in}
\end{tabular}
\end{center}
%\vspace*{-.15in}
\caption{The time-average $\rm \epsilon_{\rm ff}$ measured in 1.5-kpc wide hexagonal apertures sampling throughout 67 nearby galaxies targeted by PHANGS, as measured by \cite{leroy24}.  Measurements are plotted against the average cloud-scale velocity dispersion $\langle\sigma_{\rm mol}^{\rm cloud}\rangle$ in each aperture (\citealt{leroy24}; see \citealt{sun22,sun23} for details). Representative values for the $z\sim1$ clumps examined by \cite{dz23} are indicated by the thick gray point and bar, which depict the mean and rms of clump velocity dispersions, respectively, at the 11\% efficiency estimated for these clumps \citep{dz23}. 
Colored lines illustrate basic predictions from turbulence-regulated SF models in single-free-fall (SFF) and multi-free-fall  (MFF) scenarios: \cite{KM05} (KM05, SFF; blue), \cite{PN11} (PN11, SFF; purple, scaled down by a factor of 10) and \cite{FK12} (FK12, MFF; red, scaled down by a factor of 100).  Each model is shown at three values of the virial parameter $\alpha_{\rm vir}=2.9$, $\alpha_{\rm vir}=5.3$ and $\alpha_{\rm vir}=9.7$, the 16$^{\rm th}$, 50$^{\rm th}$ and 84$^{\rm th}$ percentiles of the plotted regions, respectively.  The linewidth of the trends increases with increasing virial parameter.  
All models adopt $\epsilon_{\rm core} =0.5$, $b=0.87$, $\phi_t=1.9$ and gas sound speed $c_s = 0.3 \, {\rm km \, s^{-1}}$ (see main text for definitions of these quantities) and, for comparison purposes, the same definition for the critical density, given by Eq. (\ref{eq:scritKM05}), estimated using the appropriate virial parameter for each line.  % measured for the plotted regions.  
 }
\label{fig:PHANGSsfe}
%\end{flushleft}
\end{figure*}
\section{Comparison with theoretical predictions}\label{sec:theorycomp}
The empirical relation between $\epsilon_{\rm ff}$ and three cloud-scale properties, $\langle\Sigma_{\rm mol}^{\rm cloud}\rangle$, $\langle\sigma_{\rm mol}^{\rm cloud}\rangle$, and $\langle\alpha_{\rm vir}^{\rm cloud}\rangle$, are presented in \cite{leroy24}.  Of the three empirical trends, $\epsilon_{\rm ff}$ vs. $\langle\sigma_{\rm mol}^{\rm cloud}\rangle$ is constructed from  independent observables, and so we take this as our diagnostic of choice in what follows.  

Figure 1 plots the kpc-scale cloud population average $\rm \epsilon_{\rm ff}^{\rm obs}$ against the average cloud-scale velocity dispersion $\langle\sigma_{\rm mol}^{\rm cloud}\rangle$ for 841 hexagonal regions sampling throughout 67 galaxies.  As noted above, the relation between $\rm \epsilon_{\rm ff}^{\rm obs}$ and $\langle\sigma_{\rm mol}^{\rm cloud}\rangle$ appears in contradiction with basic expectations from turbulence-regulated star formation models.  A summary of the models is given in the next section but here we note that, moving to regions of large velocity dispersions, elevated Mach numbers might generally be expected to create a widened distribution of densities, enhancing the amount of material that is able to collapse and form cores and stars.  It is the goal of the remainder of this work to understand the factors that lead to deviation from this scenario.

\subsection{A summary of turbulence regulation: %}\label{sec:mainfeatures}
%\subsection{
variations in the cloud-scale efficiency per free-fall time $\rm \epsilon_{\rm ff}$}\label{sec:mainfeatures}% from theory and observations}\label{sec:sfeoverview}
The two classes of models (described below) shown in Figure 1 share a number of common elements and assumptions, but differences in the way the collapse process is envisioned lead to substantially different predictions in the two cases.  Before describing those differences in $\S$ \ref{sec:sff} and \ref{sec:mff}, we first summarize the basic characteristics of turbulence-regulated star formation models.  

In theories of turbulence-regulated star formation (\citetalias{KM05}, \citetalias{PN11}, \citetalias{FK12}, \citetalias{HC11}, \citetalias{burkmocz}), the inefficiency of the star formation process taking place within cold dense molecular gas arises from two factors: 
a high density threshold for collapse $\rho_{\rm crit}$-- set by the competition between the thermal, turbulent and gravitational energy within the cloud -- and the distribution of gas densities imprinted by turbulence above and below this threshold.  

For a given density PDF, the star formation rate is derived by integrating the PDF above the critical density for collapse, yielding the so-called core formation efficiency, and then scaling this by a core-to-star efficiency $\epsilon_{\rm core}$, the efficiency with which individual cores form stars.  
Observations and simulations \citep[][\citetalias{FK12}]{enoch08, sadavoy10, alves07, andre10, matzner} suggest relatively high canonical values $\epsilon_{\rm core}\sim0.3-0.5$.  Thus the much lower 1-2\% efficiencies typical of local and extragalactic clouds are largely a consequence of the inefficiency of the core formation process.  

For convenience in comparing different observational estimates of the SFR that are sensitive to different timescales, it is common to consider the star formation efficiency per free-fall time $\epsilon_{\rm ff}$ (sometimes written SFE$_{\rm ff}$)
\begin{equation}
\epsilon_{\rm ff}=\epsilon_{\rm core}\int_{s_{\rm crit}}^{\infty}\frac{t_{\rm ff}(\rho_0)}{t_{\rm coll}(\rho)} \frac{\rho}{\rho_0} p(s)ds\label{eq:sfe1}
\end{equation}
where the free-fall time $t_{\rm ff}$ is
as given in Eq. (\ref{eq:tff}), $\rho_0$ is the mean density, $s = \ln (\rho/\rho_0)$, $s_{\rm crit} = \ln (\rho_{\rm crit}/\rho_0)$ and the PDF is normalized such that $\int_{-\infty}^{\infty}e^s p(s)ds=1$.  Here $t_{\rm ff}(\rho_0)$ is the free-fall time 
at the mean density $\rho_0$ and $t_{\rm coll}(\rho)$ is the timescale for gas at density $\rho$ to collapse to form cores. This is often written, following \citetalias{KM05}, as $t_{\rm coll}=\phi_t t_{\rm ff}$ where $\phi_t$ is a scaling factor of order unity.

In contemporary turbulence-regulated star formation models, there is some variety in the definition of the critical density $s_{\rm crit}$ \citepalias{KM05, PN11,HC11} although these tend to be similar in magnitude and behavior \citep[e.g.][\citetalias{KM05}]{burk18}.   
Here we adopt the critical density derived by \citetalias{KM05}, identified as the density at which the sonic length \citep[e.g.][]{federrath20} and the Jeans length are comparable, i.e.
\begin{equation}
\exp{(s_{\rm crit})}=\frac{\rho_{\rm crit}}{\rho_0}=\frac{\pi^2\phi_x^2}{15}\alpha_{\rm vir} M^{\frac{2}{p}-2} \approx\alpha_{\rm vir} M^{2}, \label{eq:scritKM05}
\end{equation}
where $p\sim 0.5$ is the exponent in the turbulent size-linewidth relation, $\phi_x$ is a factor of order unity, and $\alpha_{\rm vir}$ and $M$ are the cloud-scale virial parameter and Mach number, respectively.  Throughout this work we adopt the value $\phi_x$=1.12 calibrated by \citetalias{KM05} using numerical simulations.  

Most model varieties hinge on a common lognormal density distribution of the form
\begin{equation}
p(s)=\frac{1}{\sqrt{2\pi}\sigma_s}\exp{\left(\frac{-(s-s_0)^2}{2\sigma_s^2}\right)},\label{eq:LNpdf}
\end{equation}
expected from supersonic isothermal turbulence where $s_0=-1/2\sigma_s^2$ and the PDF width is set by the turbulent properties of the gas:
\begin{equation}
\sigma_s^2 = \ln \left(1 + b^2 \frac{\beta}{\beta + 1} M^2 \right),  \label{eq:sigmat}
\end{equation}
%%%%glover: You need to discuss here what you're assuming for \beta in the models presented in this paper (or mention it later, similar to what you do for b).
where $M$ is the turbulent Mach number, $b$ is the turbulent forcing parameter (ranging from $b = 1/3$ for solenoidal turbulence to $b = 1$ for irrotational turbulence), and $\beta = p_{\rm th} / p_{\rm mag}$, the ratio of the thermal and magnetic pressures \citep{federrath08,molina12,FK13}. 
In what follows we adopt $b=0.87$ and $\beta=1$ for compatibility with \citetalias{KM05}. 

Models for the density PDF that include a power-law component -- to describe the influence of self-gravity and collapse on turbulent motions and the build-up of high density material -- are also becoming more common (see e.g.~\citealt{girichidis14, burk18}, \citetalias{burkmocz},\citealt{khullar22}), as we also consider in this paper.  

The most significant differences in predictions for $\rm \epsilon_{\rm ff}$ from model to model originate with 
the manner in which collapse occurs (\citetalias{FK12}) and, in particular, whether the density structure is assumed to be in a steady-state.  In the following sections we summarize the two main scenarios and how this translates into in the rate of gravitational collapse and star formation.

\subsubsection{`Single free fall' \citetalias{KM05} and \citetalias{PN11} predictions: core formation in a single free-fall time}\label{sec:sff}
In the model of \citetalias{KM05}, collapse at all densities above $\rho_{\rm crit}$ is assumed to occur at the cloud free-fall rate, i.e. $t_{\rm coll}(\rho)=t_{\rm ff}(\rho_{\rm 0})$.  
With the density PDF effectively static in this case, as it develops only once in a cloud free-fall time, the gas at density $\rho$ that collapses to form cores also does so only once in a free-fall time. %, i.e. $t_{\rm coll}(\rho)= t_{\rm ff}(\rho_0)$. 
Thus, the so-called multi-free-fall factor $t_{\rm ff}(\rho_0)/t_{\rm coll}$ drops from the integrals and 
\begin{equation}
\epsilon_{\rm ff,SFF}=\epsilon_{\rm core}\int_{s_{\rm crit}}^{\infty}\frac{\rho}{\rho_0} p(s)ds,  \label{eq:KM05}
%SFE_{\rm ff,SFF}=\epsilon_{\rm core}\int_{s_{\rm crit}}^{\infty}e^s p(s)ds  \label{eq:KM05}
\end{equation}
which becomes
\begin{equation}
\epsilon_{\rm ff,SFF, LN}=\frac{\epsilon_{\rm core}}{2\phi_t}\left[1+{\rm erf}\left(\frac{\sigma_s^2-2s_{\rm crit}}{\sqrt{8\sigma_s^2}}\right)\right]\label{eq:sffLN}
\end{equation}
in the case of a LN density PDF, where $\phi_t$ is a factor of order unity (\citetalias{KM05}).  Throughout the remainder of this work we adopt the value $\phi_t=$1.9 calibrated by \citetalias{KM05}. 
Given our choice of $\epsilon_{\rm core}=0.5$, comparison with the predictions examined by \citetalias{FK12}, who adopt $\epsilon_{\rm core}/\phi_t=1$, would need to be scaled down by roughly a factor of 4 ($\sim$ 0.6 dex).   

Combined with the high critical density (i.e.~given by Eq.~[\ref{eq:scritKM05}]), predictions from this class of single-free-fall (SFF) turbulence-regulated theories successfully yield low, 1-2\% efficiencies, even in combination with a relatively high $\epsilon_{\rm core}\sim0.5$.  

However, as illustrated in Figure \ref{fig:PHANGSsfe}, these predictions are characterized by a weak dependence on virial parameter and Mach number (with Eq.~(\ref{eq:sffLN}) \citetalias{KM05} predict $\rm \epsilon_{\rm ff}\propto \textit{M}^{-0.37}$) that is unable to fully capture the wide range in $\epsilon_{\rm ff}$ exhibited by the observed (and simulated) cloud populations in galaxies ranging from normal star-forming disks to starbursts and high-redshift galaxies (where $\epsilon_{\rm ff}$ can reach values as high as 0.1; \citealt{usero,salim15,utomo18}; and see Figure \ref{fig:PHANGSsfe}).

This is slightly modified in the scenario envisioned by \citetalias{PN11} who assume collapse happens at the free-fall rate at the critical density.  In this case, the multi-free-fall factor $t_{\rm ff}(\rho_0)/t_{\rm coll}=t_{\rm ff}(\rho_0)/t_{\rm ff}(\rho_{\rm crit})$ is once again independent of density, but the $\epsilon_{\rm ff}$ in Eq. (\ref{eq:sffLN}) includes an additional multiplicative factor $\textrm{exp}{(s_{\rm crit}/2)}$.  
This greatly enhances the dynamic range in $\epsilon_{\rm ff}$ but, as illustrated in Figure \ref{fig:PHANGSsfe}, the predicted increase with increasing Mach number is opposite to the sense implied by extragalactic observations.     

\subsubsection{`Multi-free-fall' \citetalias{HC11}, \citetalias{FK12} predictions: core formation over multiple free-fall times}\label{sec:mff}
As argued by \citetalias{FK12}, another avenue to boost the dynamic range of predicted efficiencies is to envision the distribution of gas densities as steady-state, continuously replenished over a time $t_{\rm renew}$, as predicted for turbulent gas by \citetalias{HC11}.  As a result, collapse at the rate $1/t_{\rm coll}(\rho)\approx 1/t_{\rm ff}(\rho)$ can occur multiple times in the period $t_{\rm renew} \approx t_{\rm ff}(\rho_0)$. 

In this case, the `multi-free-fall' (MFF) factor $t_{\rm ff}(\rho_0)/t_{\rm coll}(\rho)\propto t_{\rm ff}(\rho_0)/t_{\rm ff}(\rho)$ is kept inside the integral in Eq.~(\ref{eq:sfe1}) and
\begin{equation}
\epsilon_{\rm ff,MFF, LN}=\epsilon_{\rm core}\exp \left(\frac{3}{8} \sigma_s^2 \right)\frac{1}{2}\left[1+{\rm erf}\left(\frac{\sigma_s^2-s_{\rm crit}}{\sqrt{2\sigma_s^2}}\right)\right]\label{eq:mffLN}
\end{equation}
assuming a lognormal distribution of densities (\citetalias{FK12}).  

The factor $\exp{(3\sigma_s^2/8)}$ in Eq.~(\ref{eq:mffLN}) introduces a strong increase in the predicted $\epsilon_{\rm ff}$ with increasing Mach number.  
The resulting large dynamic range allows MFF models to reach the highest star formation efficiencies observed \citep[$\sim$ 0.1;][]{FK12,salim15,utomo18,dz23} better than the \citetalias{KM05} model, which needs to invoke unphysically low Mach numbers (given the weak dependence on virial parameter and Mach number).  
This yields a qualitative match to the star formation rates of different populations, from cores to clouds and starbursts \citep{salim15}; systems that are overall more turbulent tend to have relatively high $\epsilon_{\rm ff}$, as predicted in Eq.~(\ref{eq:mffLN}).   

Similar to the \citetalias{PN11} predictions, however, the Mach number dependence at fixed $b$ and $c_s$ is opposite to the behavior exhibited by the cloud populations of normal star-forming galaxies (\citealt{leroy17,utomo18}; see Figure \ref{fig:PHANGSsfe}).
Variations in $b$ and $c_s$ (which are taken to be fixed in the right panel of Figure \ref{fig:PHANGSsfe} for illustration) might be an avenue for altering the slope of the predicted trend between $\epsilon_{\rm ff}$ and $\sigma$.  
However, with reasonable ranges in these values appropriate for the cold dense gas in PHANGS targets the predictions are not modified substantially enough to match the PHANGS measurements.  

Matching MFF predictions to observations moreover requires a significant ad hoc reduction in the normalization of the model, by one or two orders of magnitude \citep{salim15,utomo18}.  Indeed, at fixed $M$, Eq.~(\ref{eq:mffLN}) predicts values for $\epsilon_{\rm ff}$ that are a factor of 10 to 100 higher than predicted by single free-fall collapse scenario (Eq.~[\ref{eq:sffLN}]).  The factors $\phi_t$ or $\phi_x$ entering $s_{\rm crit}$ could represent reasonable paths for modifying the normalization, i.e. to apply to galactic scale turbulence that behaves differently than the forced or decaying turbulence in idealized GMC simulations.  For now we choose to adopt the values for these factors from the literature and explore other paths to reduce the normalization of the MFF predictions in $\S$ ~$\ref{sec:proposedmodifications}$.  These operate by taking into account how the galactic context of star-forming clouds impacts the output of turbulence-regulated star formation.  They thus offer an analytical description to compliment what is found in numerical simulations, namely that factors like feedback, large-scale turbulence, self-gravity and magnetic fields can reduce the output of MFF predictions to $\sim$1\% \citep[i.e.][]{federrath15,kretschmerTeyssier} or impact how closely the efficiencies produced in the simulation match the adopted sub-grid efficiency \citep{otero24}.  

\begin{figure*}[t]
%\begin{flushleft}
%\vspace*{-.15in}
\begin{center}
\begin{tabular}{cc}
\hspace*{-.6in}\includegraphics[width=0.65\linewidth]{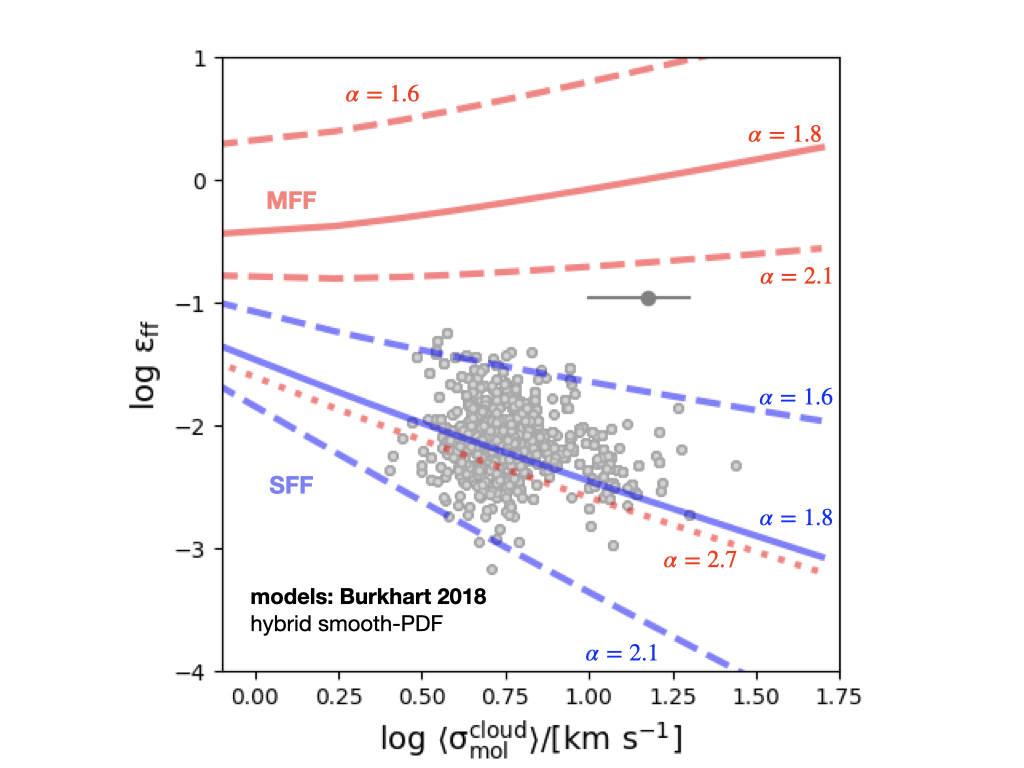}&%Bplot150nh.jpg}%modBplot150nh.jpg}%pdf}
\hspace*{-.9in}\includegraphics[width=0.65\linewidth]{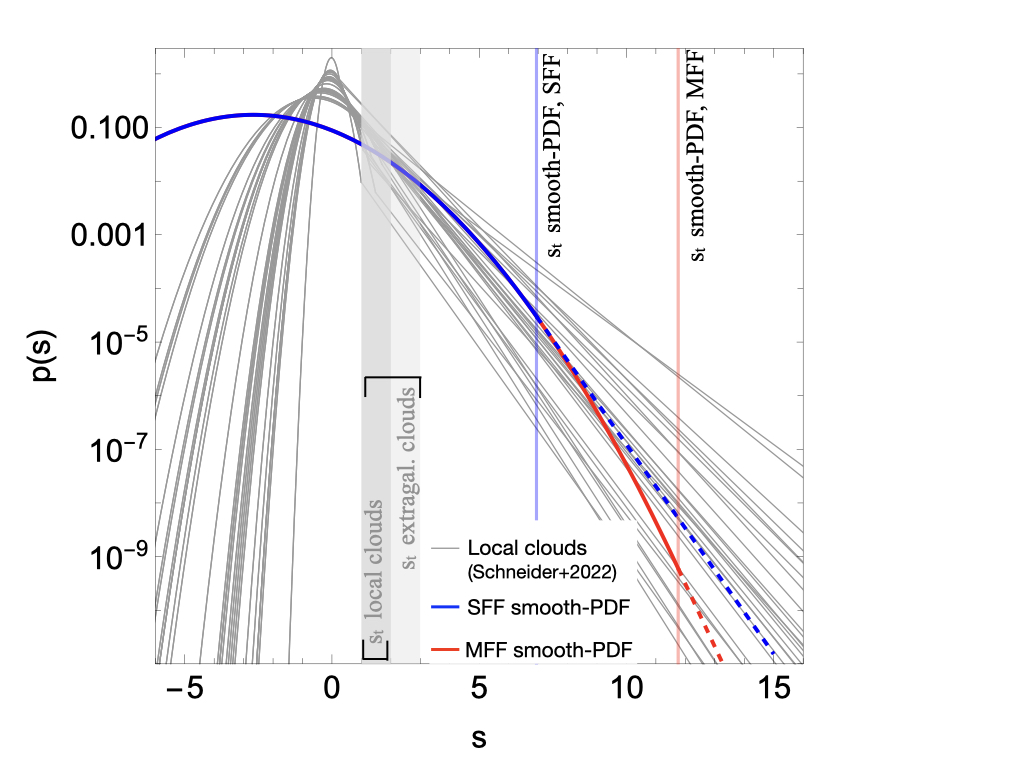}%Bplot150nh.jpg}%modBplot150nh.jpg}%pdf}
\end{tabular}
\end{center}
%\vspace*{-.15in}
\caption{(Left) Predictions for $\rm \epsilon_{\rm ff}$ from turbulence-regulated SF models with a hybrid LN+PL smooth-PDF proposed by \cite{burk18} in the single-free-fall (blue) or multi-free-fall (red) scenarios.  In these hybrid PDFs, the transition from lognormal to power-law behavior is set to the critical density, $s_t=s_{\rm crit}$, as argued by \cite{burkmocz}.  A range of power-law slopes $1.6<\alpha<2.1$ set to the range observed by \cite{Kainulainen14} and  \cite{schneider22} are indicated by the width of each band. In this example, $\epsilon_{\rm core} =0.5$, $\alpha_{\rm vir}=5$, $b=0.87$ and the sound speed $c_s = 0.3 \, {\rm km \, s^{-1}}$.  Following  \cite{burk18} we set $\phi_t=1$.  
Light gray points show the PHANGS measurements from \cite{leroy24} and the dark gray bar and point depicts the \cite{dz23} $z\sim1$ clumps, repeated from Figure \ref{fig:PHANGSsfe}. (Right) Illustration of the typical hybrid density PDFs associated with the $\rm \epsilon_{\rm ff}$ predictions in the single-free-fall (blue) and multi-free-fall (red) scenarios at the average cloud-scale velocity dispersion $\langle\sigma_c\rangle$=5 km s$^{-1}$, corresponding to $\mathcal{M}=16.7$.  The blue PDF adopts a power-law with slope $\alpha=1.8$, which is associated with $s_t$=6.9 (see eq.[\ref{eq:burkst}]; marked by the blue vertical dotted line) given the average $\sigma_c$, and yields SFF $\rm \epsilon_{\rm ff}$ predictions like those shown in blue in the left panel.  
The red PDF adopts a power-law with slope $\alpha=2.7$ from the range required to match MFF predictions to the  observed $\rm \epsilon_{\rm ff}$.  In this case the power-law starts at $s_t$=11.4 (marked by the red vertical dotted line).  Also shown are PDFs with properties matching those measured in local clouds by \cite{schneider22}.  Out of two lognormal and two power-law components identified by \cite{schneider22}, only the primary lognormal and power-law components are indicated.  These have $s_t=1-2$, marked by the narrow gray vertical band, resulting in a kinked appearance. The lighter, wider gray vertical band shows the range in $s_t$ predicted using Eg. (\ref{eq:simplerhog}) given the observed properties of the plotted regions.  }
\label{fig:Burksfe}
%\end{flushleft}
\end{figure*}

\section{Modification of MFF models}\label{sec:proposedmodifications}
Recent observational and theoretical insights suggest several avenues for adjusting the predictions of SFF and MFF models to obtain an improved match to observations. 

\subsection{An overview of the role of gas self-gravity} %A sketch of avenues for modifying models}
%Below we will sketch the modifications that are inherent to a view of self-gravitating gas.   

Already one of the leading proposals for modifying the predictions of turbulence-regulated star formation models is to add a power-law to the assumed density PDF \citep[including][]{girichidis14,burk18,meidt20,burkmocz,jaupart}.  This is a natural expectation for gas that is self-gravitating \citep{klessen, kritsuk11, Ballesteros, collins12, FK13, burk18, jaupart,donkov1,donkov2}.  It also matches the density distributions in local resolved clouds \citep{Kainulainen14, schneider15,dib20,Spilker21,schneider22}, which exhibit power-law structure \citep{Lombardi15, abreu,alves17} down to densities near the cloud edge where the HI-to-H$_2$ transition typically takes place ($\Sigma$$\sim$10-50 M$_\odot \, {\rm pc}^{-2}$).  
Although the low-density behavior of observed PDFs is debated \citep[e.g.][]{schneider15,alves17,kortgen},
 due in part to the inherent difficulty in distinguishing between low density cloud material and foreground or background contamination, the observations are compatible with the expectation that the distribution of gas densities is characterized by a LN component transitioning to a power-law component at the onset of self-gravitation somewhere near the edges of typical clouds.  

As will be described more in what follows, the PDFs designed in this work to apply under the conditions observed in extragalactic gas (relatively high Mach number and surface density) also contain a prominent PL that begins near the cloud edge and contains a large fraction of cloud material.  This prompts other changes from existing descriptions of the star formation process in which the onset of self-gravitation is either accounted for implicitly, by setting the time to rejuvenate the LN density PDF shaped by turbulence (\citetalias{HC11}, \citetalias{FK12}), or is assumed to occur at higher density in a smaller fraction of the cloud and coincide with core formation (where self-gravity just exceeds thermal and turbulent pressure; \citetalias{burkmocz}).   The next three sections discuss further motivation for these changes and how we incorporate them into turbulence-regulated star formation models.

\subsection{Lognormal PDFs with a power-law tail}\label{sec:powerlaws}

In hybrid PL+LN density PDFs, the PDF transitions to a power-law at the critical density $s_t$ for self-gravitation, i.e. 
\begin{equation}
p(s)=
\begin{cases}
 N\frac{1}{\sqrt{2\pi \sigma_s^2}}\exp{\left(\dfrac{-(s-s_0)^2}{2\sigma_s^2}\right)}& \text{$s<s_t$}\\
 N C \exp{(-\alpha s)}& \text{$s>s_t$}, \label{eq:hybridPDF}
\end{cases}
\end{equation}
where the normalization $N$ is chosen so that the PDF integrates to unity and $C$ is the amplitude of the power-law component, determined by requiring that the density PDF must be continuous \citep[see][]{burk18}.  The slope of the power-law tail $\alpha$ is sometimes interpreted in terms of the spherically symmetric radial density profile $\rho\propto r^{-k}$, where  $k=3/\alpha$ \citep{kritsuk11,girichidis14}. 

The PDF proposed by \citet[][hereafter the smooth-PDF]{burk18} is formulated in such a way that $s_t$ is also related to the index of the PL component.  Specifically, under the assumption that the density PDF must be smooth (differentiable), \cite{burk18} find that 
\begin{equation}
s_t=(\alpha-1/2)\sigma_s^2, \label{eq:burkst} 
\end{equation}
where $\alpha$ is the power-law index and  $\sigma_s^2$ is given by Equation~(\ref{eq:sigmat}).

In the context of turbulence-regulated star formation, adding a PL tail to the density PDF has a number of practical advantages. 
As shown by \cite{burk18}, power-laws are able to increase the dynamic range of predictions at fixed Mach number.  They also reduce the output of the models by far enough that \citetalias{burkmocz} propose that this can remove the need to artificially lower the normalization of MFF predictions (i.e. through $\epsilon_{\rm core}$) in order to match observations \citep{salim15, utomo18}.  

Moreover, with the functionality of Eq. (\ref{eq:burkst}) built in, variations in the power-law part of the PDF introduce strong changes in the efficiency per free-fall time, especially when the transition to self-gravitation is used as the critical threshold for collapse and core formation as argued by \citetalias{burkmocz}.  In such a scenario, as $s_t$ is lowered, the power-law slope becomes shallower and the self-gravitating fraction in the cloud increases, with the result that $\epsilon_{\rm ff}$ increases with decreasing $\alpha$ (increasing $k$).  

\subsubsection{Efficiencies predicted with smooth LN+PL PDFs}%Comparison to observations}
Figure \ref{fig:Burksfe} shows the $\rm \epsilon_{\rm ff}$ predicted in single and multi-free-fall scenarios with a hybrid smooth-PDF where $s_t=s_{\rm crit}$ (\citetalias{burkmocz}) and the power-law slope is fixed to fall within the range $1.5<\alpha<2.5$, in line with theoretical predictions \citep{Penston69,shu77,kritsuk11,girichidis14,khullar22,donkov1,donkov2} and the structure in resolved clouds \citep[e.g.][]{Kainulainen14,Spilker21,schneider22}.  
The predictions here (as in Fig.~\ref{fig:PHANGSsfe}) assume a thermal sound speed of 0.3 km s$^{-1}$, $\epsilon_{\rm core}=0.5$, and turbulent forcing parameter $b=0.87$.   In the \cite{burk18} formulation, $\phi_t =1$ and we adopt that here when using any LN+PL PDF.  

There are clear differences in the predictions from the two classes of models, with the SFF scenario standing out as providing the closer match to observations.  As emphasized by \cite{burk18} LN+PL predictions can easily reach $\epsilon_{\rm ff}=1-2$\% even when $\epsilon_{\rm core}\approx 0.5-1$.  
However, here in Figure \ref{fig:Burksfe} this has arguably less to do with the addition of the PL (in contrast to the speculation of those authors) than with the SFF nature of the prediction.  Like SFF predictions generally, the values plotted in Fig.~\ref{fig:Burksfe} may pass through the PHANGS measurements, but they fall shy of some of the highest observed values $\rm \epsilon_{\rm ff}\sim 10\%$ 
 \citep{salim15, dz23}.  

The situation is slightly improved in the multi-free-fall scenario, with $\rm \epsilon_{\rm ff}$ easily reaching $\gtrsim$0.1, but still the lowest values for $\rm \epsilon_{ff}$ are out of reach for models with $\alpha$ in the observed range $1.5<\alpha<2.5$.  Adding a power-law tail to the PDF still does not solve the issue for PLs with such slopes.  
To match the measured $\epsilon_{\rm ff}$ in this scenario with relatively high Mach numbers ($8\gtrsim \mathcal{M}\gtrsim 100$) would require significantly steeper power-law components with $2.4<\alpha<3$ (fitted in $\S$ \ref{sec:obsTests}) that are incompatible with the observed clouds or with theory.   

\subsection{The onset of self-gravitation}
A less readily apparent but nonetheless critical issue with either (MFF or SFF) variety adopting a smooth hybrid PDF is what the transition density between the LN and PL components implies about where gas becomes self-gravitating and how much material is contained at the highest densities.    

For gas with Mach numbers as high as observed in extragalactic targets ($8\lesssim \mathcal{M}\lesssim 100$ in Figure 1), the $s_t$ associated with $1.5<\alpha<2.5$ falls in the range $5-11$.  These high $s_t$ are compellingly close to the critical density for collapse in Eq. (\ref{eq:scritKM05}), as highlighted (and leveraged) by \citetalias{burkmocz}\footnote{Conservatively assuming $\alpha_{vir}\sim 1$, $4<s_{\rm crit}<8$ for $5\gtrsim \mathcal{M}\gtrsim 8$.  }.  However, such high values unavoidably leave an extended, broad lognormal component throughout the cloud.  This places 
substantially more mass in intermediate and high density material than when the power-law begins, e.g., nearer to the cloud edge, as observed in local clouds.  Although we will make a fuller study in $\S$ \ref{sec:results}, here we note that clouds with hybrid smooth-PDFs containing a small $\alpha\sim1.6$ PL tail have typically 5-30\% of their mass above a density threshold of 10$^3$ cm$^{-3}$ \citep[the effective critical density for HCN, corresponding roughly to the minimum density of the gas responsible for most of the HCN emission from nearby clouds;][]{leroydense,neumann23}.  Extragalactic dense gas fractions, on the other hand, are normally observed in the range 1-15\% \citep{gallagher18,neumann23}. 

A number of arguments favor a lower transition to self-gravitation, including the observed properties of clouds both in the MW and extragalactic targets.  Extragalactic clouds probed over a large range of galactic environments, for one, often show an excess of kinetic energy on the cloud scale. This leads to a super-virial dynamical state, as measured by comparing the kinetic energy in the gas to its self-gravity.  For the regions under study in Figure 1, $2\lesssim\alpha_{\rm vir}\lesssim 10$ \citep[][and as noted in the caption to Figure 1]{leroy24}.  With a few assumptions, those virial levels can be used to place constraints on the densities needed for gas-self gravity to overcome the turbulent energy.  
Consider, for illustration purposes, a spherically symmetric cloud with an internal power-law density profile $\rho\propto r^{-k}$ where $k=3/\alpha$ \citep{kritsuk11,girichidis14}.  Let the turbulence in the cloud obey the relation $\sigma^2=\sigma_0^2(r/r_0)^l$, where $\sigma_0$ is the gas turbulent velocity dispersion on the outer scale $r_0$ and the power-law index $l\sim1$ characteristic of the turbulence \citep[i.e.][]{mckeeOstriker07,heyerdame}.  With these assumptions, the virial parameter at any radius $r$ in the cloud is related to the virial parameter $\alpha_{\rm vir,0}$ at radius $r_0$, i.e.  
\begin{equation}
\alpha_{\rm vir}  =\alpha_{\rm vir,0}\left(\frac{r}{r_0}\right)^{k+l-2}.
\end{equation}
According to this relation, the virial level $\alpha_{\rm vir,0}>1$ observed for any cloud on outer scale $r_0$ informs the density $\rho_G$ needed for gas to become self-gravitating.  Letting $\alpha_{\rm vir}=2$ mark the threshold for self-gravitation, 
then
\begin{equation}
\left(\frac{\rho_G}{\rho_0}\right)=\left(\frac{\alpha_{\rm vir,0}}{\alpha_{\rm vir}=2}\right)^{\frac{k}{k+l-2}}.\label{eq:simplerhog}
\end{equation}
Assuming $l=1$ and letting $k=1.5$ for illustration, we estimate $1.2\lesssim\ln \rho_G/\rho_0\lesssim 5$ for gas with $3\lesssim\alpha_{\rm vir}\lesssim 10$ measured in PHANGS targets \citep[][]{leroy24}.  This lowers to $1\lesssim\ln \rho_G/\rho_0\lesssim3$ in the case that $k=2$.

These values are not only substantially lower than required to match hybrid smooth PDFs to the observed gas properties, they are also much closer to the edge of the cloud, more reminiscent of where the PL starts in local clouds \citep{schneider22}   This is arguably consistent with the expectation that the self-gravity of the gas must start being non-negligible somewhere near the cloud edge or the gas would not have started to clump and form the cloud or shape its internal structure, even with the assistance of turbulence.  In other words, the defining of the cloud implies that self-gravitation must be happening somewhere not too far from its `edge'.   

Later in $\S$ \ref{sec:M20bottleneck}, we invoke a model that relates the turbulent motions observed on the cloud scale -- and the super-virial appearance of the gas -- to motion in the galactic potential developing under galactic gravitational forces \citep{meidt18,meidt20}.  We then use that model to propose a specific threshold for the onset of self-gravitation $\rho_G$.  The alternative to using this model for turbulent gas motions would be to use the observed gas motions directly.   We attempt to leave our approach generic enough to accommodate such a choice, leaving the source of the turbulent motion unspecified, or to introduce a different model for the turbulence altogether.   However, since our model provides a description for the observed systematic variation in cloud velocity dispersions and virial state \citep{meidt18}, we find it a useful framework for interpreting structural variations and how they may be expected to vary with galactic environment (as explored later in $\S$ \ref{sec:results}).    

\subsubsection{A broad PL tail in a kinked LN+PL PDF?}\label{sec:powerlawchange}
As underlined by the comparison in Figure 2, matching the observed $\epsilon_{\rm ff}$ with a hybrid smooth-PDF in a MFF scenario is difficult when the power-law is as shallow ($\alpha\sim 1.5-2.5$) as implied by observed clouds and as expected from theory \citep{larson69, Penston69, shu77, kritsuk11, girichidis14,jaupart, khullar22}.  
A relaxation of the differentiability criterion\footnote{Whereas the distribution function, or cumulative density function, needs to be continuous and differentiable, the density PDF does not.} adopted by \cite{burk18}, however, allows for 
a more flexible and physical LN+PL model that can match observed PDFs and expectations for the onset of self-gravitation while keeping the predicted star formation efficiency low.  

When the differentiability criterion is removed, the hybrid PL+LN PDF has the same normalization as derived by \cite{burk18}, but is no longer restricted to a particular $s_t$ for a given $\alpha$ (or vice versa).  This gives the model the leverage to match observed star formation rates with realistic values for $\alpha$ in high Mach number gas all while $s_t$ is lowered (i.e. below the values required in Figure 2).  The resulting PDF has a kink where the PDF transitions from lognormal to power-law, as is often observed in clouds \citep[e.g. Figure 2;][]{Kainulainen14, Spilker21,schneider22}.  
A major advantage to this sort of non-smooth, kinked hybrid PDF is that it can place significantly less mass at intemediate-to-high densities compared to equal-mass lognormal and smooth-hybrid counterparts (discussed further in $\S$ \ref{sec:results}), yielding a reduced core formation rate in MFF scenarios.  Thus, even in gas with relatively high Mach number, PDFs can yield low enough MFF efficiencies to match observed cloud populations when the LN to PL transition is shifted down and the PDF contains an extended, broad PL.  

Inherent to this modification, we propose that the transition to self-gravitation does not necessarily mark a transition to free-fall collapse, or even coincide with critical density for core formation, i.e.~$s_t\neq s_{\rm crit}$.  This is based on the idea that self-gravity is capable of shaping the PL even if it does not dominate the energy associated with thermal and turbulent pressure or magnetic fields on some scales. 
In the presence of these factors, theory and simulations suggest that the action of gravity is to collect material and enhance density contrasts, at most engaging in a slowed or delayed collapse \citep[][]{girichidis14,xuLazarian, jaupart,khullar22}.  The result is recorded in the formation of the power-law portion of the PDF with a slope that approaches, but does not reach, $\alpha=3/2$ characteristic of (pressure free) free-fall collapse.  

In this light, we reserve core formation specifically to the gas with self-gravity large enough to dominate turbulent and thermal energy, located by the critical density $s_{\rm crit}$. An even higher density threshold within the core boundary presumably marks where self-gravity becomes so strong that gas is able to undergo free-fall collapse to form stars.  In that gas, we would expect a second PL PDF slope that most closely approaches $\alpha=1.5$ \citep[e.g.][]{girichidis14, jaupart, khullar22}.  The onset of this second slope is arguably a factor that impacts the value of the core-to-star efficiency adopted in this work.  Since we adopt empirical calibrations of $\epsilon_{\rm core}$ and restrict ourselves to describing the phase of core formation, the first, `non-free-fall' power-law slope is expected to be most relevant.  

\subsection{Finite replenishment of density structure}\label{sec:trenew}
By envisioning the gas as self-gravitating and adding a PL to the PDF we are also prompted to reconsider how gas structure builds and renews.  Our preference to place material in a broad power-law tail \citep[so that the PL slope is consistent with the structure in resolved clouds;][]{Kainulainen14} 
is conceptually consistent with the idea of pervasive hierarchical gravitational collapse and turbulence that acts to continuously replenish the density structure in clouds over the course of a cloud free-fall time (\citetalias{HC11}, \citetalias{FK12},  \citealt{girichidis14, appel22}).  That idea leads to the expectation that the time to rejunevate self-gravitating structures is the timescale over which the PDF is renewed.  This is the central idea behind the multi-free-fall scenario, which we select as our fiducial model in what follows.\footnote{With their choice of $s_t=s_{\rm crit}$, \citetalias{burkmocz}, in contrast, implement what is essentially a single-free-fall scenario to predict SFEs.  

Now that we allow self-gravitation to explicitly shape the PDF and add a PL, however, it also becomes relevant to account for the time it takes for structure to rejuvenate in comparison to the cloud's lifetime $\tau\approx t_{\rm ff}(\rho_0)$.  
}
This becomes important for clouds that are not virialized on the cloud scale and instead contain an excess of turbulent energy.  For these clouds, structure build-up in the cloud at the outset is through turbulence, i.e. the turbulence crossing time $t_{\rm turb}=R_c/\sigma$ on the cloud scale $R_c$ is shorter than $t_{\rm ff,0}$.  The rejuvenation time only switches to the free-fall time at higher densities within the cloud, where self-gravity dominates (\citetalias{HC11}). By construction, precisely at the self-gravitating threshold the crossing time and the free-fall time are equal.  
Thus the free-fall time at the self-gravitating threshold is a measure of the time for self-gravity to renew the cloud's PDF.  

With this view in mind, any factors that weaken self-gravity on the cloud scale are factors that shorten the duration of PDF renewal and thus star formation.  In this work (later in $\S$ \ref{sec:M20bottleneck}) we consider self-gravity in relation to the background galaxy potential, which coordinates turbulent motions with an energy that can exceed self-gravity towards the outer cloud scale. 
The weakness of self-gravity there acts to halt gravitational collapse below a critical density, reducing the supply of collapsing material that can replenish the high density PDF.  This stops the core (and ultimately star) formation process from proceeding continuously as assumed in MFF models.  

A number of factors are also capable of preventing collapse or inhibiting mass transfer from the lognormal to the power-law part of the PDF, such as magnetic fields \citep{girichidis14} or feedback-driven outflows \citep{appel22}.  
As a second modification to conventional turbulence-regulated SF models, we therefore propose 
to loosen the steady-state assumption that is inherent in typical MFF scenarios.  In practice we do this by limiting the duration of PDF renewal to some time $t_{\rm renew}$ below the cloud free-fall time.  Changes in the PDF over the duration of PDF renewal are ignored for the present (i.e the slope of the density power-law is assumed to be constant in time). 
However, in what follows, we write the density PDF as an explicit function of time, so that models for time variation, e.g.~capturing an accelerating $\rm \epsilon_{\rm ff}$ \citep{Hartmann, murray, lee15, caldwell18}, could be incorporated in future applications.  Later in $\S$ \ref{sec:discussion} we briefly consider evolution in $\alpha$.   

For transparency, in our limited-replenishment scenario we write $\epsilon_{\rm ff}$ for a 
cloud observed at any time $t_{\rm obs}$ since the start of the star formation process as 
\begin{equation}
\epsilon_{\rm ff}=\frac{\epsilon_{\rm core}}{\phi_t}\frac{1}{t_{\rm obs}}\int_0^{t_{\rm obs}}\left[\int_{s_{\rm crit}}^{\infty}\frac{t_{\rm ff}(\rho_0)}{t_{\rm ff}(\rho)} \frac{\rho}{\rho_0} p(s,t)ds\right]dt, \label{eq:sfe1stp} 
\end{equation}
where the PDF has been normalized such that $\int_{-\infty}^{\infty}e^s p(s)ds=1$ and the term in square brackets is the cloud free-fall time multiplied by the core formation rate.  

In the event of finite PDF renewal with duration $t_{\rm renew}$, the integrand in the square brackets is non-zero only up until $t_{\rm stop}=t_{\rm renew}$, allowing core formation to proceed until $t_{\rm stop}$.     
In this work we wish to examine scenarios in which 
$t_{\rm stop}$ does not necessarily extend beyond $t_{\rm obs}$.  This is different than the typical approach, in which the core formation process effectively spans a single cloud free-fall time $t_{\rm ff}(\rho_0)$ (assuming that the cloud's lifetime is the free fall time) and this $t_{\rm ff}(\rho_0)$ is also assumed to be roughly equal to or exceed $t_{\rm obs}$, so that $t_{\rm stop}=t_{\rm ff}(\rho_0)\gtrsim t_{\rm obs}$.  For that case, the $\epsilon_{\rm ff}$ is determined entirely by the term in the square brackets (as in the previous section). For convenience, Table \ref{tab:timescales} collects the definitions of the various timescales used in this work.  

With core formation stopped by a cessation of PDF renewal before a full cloud free-fall time has elapsed, Eq.~(\ref{eq:sfe1stp}) suggests
\begin{equation}
\epsilon_{\rm ff}=\frac{t_{\rm stop}}{t_{\rm obs}}\epsilon_{\rm ff, steady},
\label{eq:sfe2}
\end{equation}
where $\rm \epsilon_{\rm ff, steady}$ is the steady-state efficiency per free-fall time that follows from multiplying $\epsilon_{\rm core}$ by the term in the square brackets in Eq.~(\ref{eq:sfe1stp}) calculated over a cloud free-fall time (i.e. with no stop to core formation).   Although we have kept Eq. (\ref{eq:sfe2}) written in terms of a generic $t_{\rm stop}$, in our preferred view $t_{\rm stop}$ is meant to be set by the free-fall time where the gas becomes self-gravitating.  

Equation (\ref{eq:sfe1stp}) can be written more generally in terms of the time when the core formation process begins $t_{\rm start}$ and the visibility timescale of the star formation tracer $t_{\rm sf}$.  For star formation traced by YSOs in galactic clouds, for example, this would be the $\sim$ 0.5 Myr duration of the protostellar phase.  
The visibility timescale is typically longer when the time-averaged star formation rate within an (extragalactic) cloud population is constructed from kpc-scale observations of extragalactic tracers like H$_{\alpha}$, FUV and 24 $\mu m$ emission, following the approach developed by \cite{leroy17} and implemented by \cite{sun23,leroy24} (yielding the measurements studied here). As we discuss later in $\S$ \ref{sec:tobs}, in this scenario, when the visibility timescale approaches or exceeds the typical cloud lifetime $t_{\rm life}$, then for any individual cloud the minimum $t_{\rm sf}=t_{\rm obs}\approx t_{\rm life}$, or roughly a cloud free-fall time \citep{Chevance20}. 

Clearly, the impact of finite $t_{\rm stop}$ will be recognizable as long as $t_{\rm obs}>t_{\rm stop}$.  This correction is therefore potentially most relevant for comparing to extragalactic efficiencies measured with long-timescale star formation tracers, i.e. if a steady-state cannot be assumed for the duration of a cloud free-fall time.  

\begin{table*}
\begin{center}
\caption{Definitions of timescales considered in this work}\label{tab:timescales}
\begin{threeparttable}
\begin{tabular}{p{0.1\linewidth} p{0.8\linewidth}}
%\begin{array}{rcc}
Symbol&Description\\
\hline
$t_{\rm ff}(\rho_0)$ & the cloud free-fall time, or the free-fall time for an object with mean density $\rho_0$\\
$t_{\rm ff}(\rho)$& the free-fall time at density $\rho$ \\
$t_{\rm coll}(\rho)$& the time for gas at density $\rho$ to collapse and form cores, often written as $t_{\rm coll}=t_{\rm ff}$, where $\phi_t$ is a scaling factor of order unity.\\
$t_{\rm renew}$& the length of time that a cloud's density PDF is renewed relative to the moment its density PDF was set at $t=0$, or the duration of PDF renewal\\
$t_{\rm start}$& the moment core formation begins in the cloud relative to $t=0$\\
$t_{\rm sf}$& the visibility timescale of any given star formation tracer \\
$t_{\rm life}$& the cloud lifetime  \\
$t_{\rm obs}$& the time when a cloud is observed relative to $t=0$.  For clouds observed with long visibility (extragalactic) star formation tracers, $t_{\rm obs}\approx t_{\rm life}$ \\
$t_{\rm stop}$& the time when a cloud's PDF renewal stops relative to $t=0$.  For long-timescale star formation tracers, $t_{\rm obs}>t_{\rm stop}$.  \\
$t_{\rm no~cl}$& the time between cloud formation events \\
$t_{\rm no~sf}$& the time clouds spend in an inert phase, between star formation events \\
\hline
\end{tabular}
\end{threeparttable}

\end{center}
%{Table}
%\label{tab:glossary}
\end{table*}

\section{A physical model for the proposed modifications: the role of galactic environment}\label{sec:M20bottleneck}
One of the main motivations for introducing both of the proposed changes to MFF predictions in the previous section is to model the influence of galactic environment on star-forming clouds. 
In turbulence-regulated star formation models, the key bottleneck to star formation is the gravitational collapse of gas to form cores, set by the competition between gravity and the energy in thermal and turbulent motions. In \cite{meidt20}, we hypothesized that a secondary bottleneck acts within clouds, taking place on larger scales and at lower densities.  
This gas is kinematically coupled to the host galaxy potential and appears super-virial \citep[e.g.][]{meidt18}.  As a result of the weakness of self-gravity (relative to the total gravitational potential of the galaxy at large), gravitational collapse occurs on timescales that are much longer than the cloud free-fall time.  This effectively inhibits self-gravitation and impedes collapse for a portion of the cloud.  Only when the gas decouples from the galaxy and achieves virial balance is collapse possible.  

With this picture in mind, we let the density threshold for gas to decouple from the galactic potential -- also referred to as the threshold for self-gravitation $\rho_{G}$ -- both mark the density where the density PDF transitions from LN to PL in the hybrid density PDF described in $\S$ \ref{sec:powerlaws} and set the timescale for PDF replenishment following the formalism presented in $\S$ \ref{sec:trenew}.    Given the typical conditions observed in extragalactic star forming disks, the transition to self-gravitation $\rho_G$ implied in this scenario places a large fraction of cloud material in a PL component, impacting the efficiency predicted in MFF scenarios as discussed in $\S$ \ref{sec:powerlawchange}.  

Again, although the focus here is on the bottleneck model, alternatives can be easily tested using the generic framework presented in $\S\S$ \ref{sec:powerlaws} and \ref{sec:trenew}.   
Indeed, the bottleneck model is in the family of theories in which self-gravitation decreases with increasing virial parameter \citep{klessen,padoan16,jaupart}.\footnote{In the  \cite{meidt18,meidt20} picture,  turbulent motions in the gas, and thus its virial state, are modeled as the equilibrium gas response in the presence of the external host galaxy potential (together with self-gravity).  The larger the galactic component, the more weakly self-gravitating and apparently `super-virial' the gas becomes, reducing its ability to collapse and form stars. } 
Most generically, we propose parameterizing the effect of super-virial motions by using $t_{\rm stop}$ and setting it to the collapse time in the portion of the gas that reaches a virial (self-gravitating) state.  This makes it possible to implement alternative models for how other (non-gravitational) processes in the gas, such as SNe feedback  (\citetalias{PN11}) or magnetic fields \citep{federrath15}, impact gas dynamical (virial) state and star formation.  

Likewise, the model presented below can be easily modified using alternative prescriptions for $\rho_{\rm G}$.  For the present work, the focus is on the threshold for dynamical decoupling from the galactic potential, but star formation efficiencies yielded by PDFs tailored to any number of thresholds can be easily tested.  

\subsection{The critical density for self-gravitation: where gas motions decouple from the host galaxy}\label{sec:decouplethresh}
\subsubsection{Practical considerations}\label{sec:potentials}
In the \cite{meidt18,meidt20} picture, the threshold density for gas motions to decouple from the galactic potential  
can be identified by examining the balance between gas self-gravity and the external potential.  
While it may be possible to compute these potentials numerically, e.g. from an observed or simulated distribution, for practical purposes we seek an analytical threshold that can be easily incorporated into the calculations in Section~\ref{sec:mainfeatures}.  The easiest way to obtain the threshold is thus to make some simplifying assumptions.   
In particular, within any given local patch of gas, we adopt a triaxial geometry and seek the threshold as the density at some triaxial boundary inside of which the gas self-gravity (counting up all the mass within the triaxial region) exceeds the external potential.  

For calculating the balance of potential and/or kinetic energies on the cloud scale, how the mass is arranged within the boundary is less important than the total mass inside that boundary.  For simplicity, though, we will assume that mass distributions within each triaxial region are cloud-like and fall away in a power-law fashion from some central reference point (to pin the integrals for the potential energies) such that $\rho\propto r^{-k}$.   
Of course, real clouds are not close to spherically symmetric or even triaxial nor do they contain only a single overdensity.  It is therefore worth emphasizing that this assumption is only invoked for assessing the potential energy near the outer part of the cloud and  has little connection to the internal density structure assumed when predicting the star formation rate.  
It is therefore not inconsistent with a more realistic, complex arrangement of mass towards higher density (i.e. with multiple peaks).  Note, too, that although the cloud objects envisioned in this scenario would themselves not be filamentary, they could be embedded in larger-scale structures with a filamentary quality \citep[e.g.][]{smith20,neralwar22,meidt23}.

The assumption that material within the cloud falls away like $\rho\propto r^{-k}$ has been fairly common in the literature \citep[e.g.][and others]{bertoldimckee,heyerdame} and does a fairly good job of matching observed cloud-scale gas motions \citep{heyer,hughes13,rosolowsky21} under the assumption of virial equilibrium.  
With this density arrangement, we write the 1D velocity dispersion
\begin{equation}
\sigma_{\rm sg}=\sqrt{2\pi (a_k/5) G\Sigma R}\label{eq:vdisp}
\end{equation} 
in terms of the gas surface density $\Sigma$ at radius $R$ in the cloud and 
the geometric factor 
\begin{equation}
a_k=\frac{(1-k/3)}{(1-2k/5)}
\end{equation}  
\citep{bertoldimckee}.  We have chosen the velocity dispersion to reflect a direct equivalence between kinetic and potential energy, rather than virial balance, thus introducing the factor of 2 in Eq. (\ref{eq:vdisp}).    

In this arrangement, we also can also write a simple expression for the motion in the galactic potential at radius $R$ within the cloud \citep{meidt18,meidt20}, 
\begin{equation}
\sigma_{\rm gal}\approx\sqrt{(b_k)} \kappa_e R,
\end{equation} 
where
\begin{equation}
\kappa_e\approx\sqrt{(\kappa^2+2\Omega^2+\nu^2)/3}
\end{equation}
 in the cloud frame, in terms of the circular frequency $\Omega$ and the radial and vertical epicyclic frequencies $\kappa$ and $\nu$ that measure the strength of galactic gravitational forces in the radial and vertical direction, respectively (see also Appendx \ref{sec:appendixgamma}).  Here $b_k$ is a geometric factor of order unity derived in Appendix \ref{sec:appendixbk}.  

With these assumptions we write the ratio between self-gravity and the background potential 
as 
\begin{eqnarray}
\gamma(R,k)&=&\left(\frac{\Phi_{\rm cloud}}{\Phi_{\rm gal}}\right)^{1/2}=\frac{\sqrt{2\pi (a_k/5) G\Sigma R}}{\sigma_{\rm gal}(R,k)}\\&=&\frac{\sqrt{2\pi (a_k/5) G\rho}}{\sqrt{b_k}\kappa_e}.\label{eq:gamma1}
\end{eqnarray}
The value of this ratio specifically at density $\rho_0$ at the cloud edge is referred to as $\gamma_0$.  In the case that the velocity dispersion in the gas reflects the quadrature sum of the velocity dispersions associated with self-gravity and the background potential \citep[i.e. the two potentials are approximately separable;][]{meidt18} then at minimum, neglecting any non-gravitational motions, 
\begin{equation}
\alpha_{\rm vir}=2+\frac{2}{\gamma(R,k)}. \label{eq:alphavirgamma}
\end{equation}

The minimum value of $\gamma$ required for collapse approaching the free-fall rate in the absence of pressure indicates where self-gravity can be expected to dominate over the background potential.  \citealt{meidt18} estimate that this coincides with $\gamma(k=0)\approx2.5$, solving the equation of motion for the collapse of a spherical shell in a pressure-free uniform density sphere in the presence of the local galactic potential.  To distinguish this criterion from the condition where self-gravity overcomes sources of (thermal, turbulent or magnetic) pressure in the cloud in order to undergo free-fall collapse, we will call rename it $\gamma_G$.   We thus designate $\gamma_{\rm G}(k=0)\approx2.5$ as the uniform-density criterion for the onset of self-gravitation.  This corresponds to $\alpha_{\rm vir}\sim2.8$ according to Eq. (\ref{eq:alphavirgamma}).  

The self-gravitation condition can be adjusted for any arbitrary (non-uniform) density distribution using the definition in Eq. (\ref{eq:gamma1}), which implies that\footnote{This is in contrast to the scaling $a_k/5$ erroneously suggested earlier by \cite{meidt18}.}
\begin{equation}
\gamma_{\rm G}(k)=\gamma_{\rm G}(k=0)\left(\frac{a_k b_0}{a_0 b_k}\right)^{1/2}.\label{eq:gamma_0}
\end{equation}
For $k=2$, for example, we have a more easily passed threshold $\gamma_{\rm G}(k=2)=\sqrt{3}\gamma_{\rm G}(k=0)$.  % for collapse.  

A further rearrangement of Eq. (\ref{eq:gamma1}) yields the threshold density  
\begin{equation}
\rho_{\rm G}=\gamma_{\rm G}^2(k) \frac{\kappa_e^2b_k}{2\pi (a_k/5) G},\label{eq:rhocoll}
\end{equation} 
where self-gravity dominates the galactic potential, in terms of $\gamma_{\rm G}(k)$ for any $k$.  
We note that the threshold here is comparable to the (radially-varying) mid-plane density of the background host galaxy $\rho_{\rm gal}$ \citep{meidt20}, i.e. 
$\rho_{\rm G}\approx 2.5^2\nu^2/(\pi G)\approx\rho_{\rm gal}$.  

In practice, the self-gravitation threshold in a region with density $\rho_0$ can be assigned independently of the cloud internal structure.  Rewriting $\kappa_e$ as $\sqrt{2(a_k/5)\pi G\rho_0}/(\sqrt{b_k}\gamma_0)$, Eq. (\ref{eq:rhocoll}) can be expressed as 
\begin{equation}
\rho_{\rm G}=\rho_0\left(\frac{\gamma_{\rm G}(k)}{\gamma_0(k)}\right)^2=\rho_0\left(\frac{\gamma_{\rm G}(k=0)}{\gamma_0(k=0)}\right)^2. \label{eq:collthresh}
\end{equation} 
The equivalence of the two terms on right hand side follows from the shared dependence of  $\gamma_{\rm G}(k)$ and $\gamma_0(k)$ on $k$.  

Even though $\gamma_{\rm G}$ and $\rho_{\rm G}$ depend on $k$, the ratio $\rho_{\rm G}/\rho_0$ is independent of internal structure since $\rho_0$ also varies with $k$.   
This 
leads to a favorable disconnect between the adopted threshold density and the structure of higher density star-forming gas and underscores the relative unimportance of the precise details of our envisioned cloud geometry on our predicted star formation efficiencies.   Indeed, in the scenarios of greatest relevance here, $\epsilon_{\rm ff}$ is almost insensitive to the adopted threshold density.    
For typical molecular gas in nearby galaxies, $\gamma_0\approx 0.5-2$ on the cloud scale \citep[][see also Figure \ref{fig:gammas}]{meidt20}.  According to Eq. (\ref{eq:collthresh}), this makes $\ln \rho_{\rm G}/\rho_0\approx 0.4-3.2$, placing $\rho_{\rm G}$ towards the outer edge.  Taking the typical observed $\langle\Sigma_{\rm mol}^{\rm cloud}\rangle\sim 10^{1.8}$~M$_{\rm \odot}$~pc$^{-2}$\citep{leroy24} and our adopted $h=100$~pc, the average cloud volume density $\langle\rho_{\rm mol}^{\rm cloud}\rangle=\langle\Sigma_{\rm mol}^{\rm cloud}\rangle/(2h)\sim 0.5$~M$_{\rm \odot}$~pc$^{-3}$.  Using this as our estimate of $\rho_0$ in Eq. (\ref{eq:collthresh}) we find $\rho_{\rm G}\sim3$~M$_{\rm \odot}$~pc$^{-3}$ (or $n\sim 100$~cm$^{-3}$).  The implication of this value of $\rho_G$ is that most of the cloud material sits in a broad power-law tail \citep[see also][]{alves17}.  As long as the power-law tail is dominant like this, in practice the precise location of $\rho_{\rm G}$ is less consequential for the $\epsilon_{\rm ff}$ than the slope of the power-law or the critical density.  

\subsubsection{A characteristically broad power-law tail: less mass at intermediate-to-high densities than in lognormal PDFs}\label{sec:broadPL}
Alongside a restriction to PDF replenishment ($\S$ \ref{sec:tstopmodel}), one of the defining characteristics of the secondary galactic bottleneck is the expectation of a broad power-law tail in the density PDF.  Considering that this is arguably a general scenario, it is worth emphasizing the number of consequences PLs have for the predicted efficiency of star formation.   We illustrate these here 
using a simple power law PDF of the form
\begin{equation}
p(s)=C_{\rm PL}\exp{(-\alpha s)},
\end{equation}
where the normalization $C_{\rm PL}$ is chosen so that $\int_{0}^{\infty}e^s p(s)=1$ (integrating out to the cloud edge at $\rho_0$).  
 
Compared to an equal-mass lognormal counterpart, power-law PDFs (with exceptions) are characterized by less mass at intermediate-to-high densities, essentially shifting the mass in that regime to both lower and higher densities (see, e.g., the inset in the right panel of Fig. \ref{fig:newmodelsfe}).  This has the practical consequence of reducing the output of turbulence-regulated star formation models.  
In the multi-free-fall collapse scenario, 
\begin{equation}
\epsilon_{\rm ff,MFF,PL}= \epsilon_{\rm core}\left[\frac{\alpha-1}{\alpha-3/2}\right] \left(\frac{\rho_{\rm crit}}{\rho_0}\right)^{(3/2-\alpha)},\label{eq:mffPL}
\end{equation} 
where that the normalization factor in square brackets is equivalent to the geometric factor $(2/3)(3-k)/(2-k)$ derived by \cite{tan06} \citep[and used by][]{meidt20} for the spherically symmetric density profile $\rho\propto r^{-k}$.  
Likewise, in the original single free-fall collapse model envisioned by \citetalias{KM05}, 
\begin{equation}
\epsilon_{\rm ff,SFF,PL}= \epsilon_{\rm core}\left(\frac{\rho_{\rm crit}}{\rho_0}\right)^{(1-\alpha)}.\label{eq:sffPL}
\end{equation}

For PLs down to $\alpha\sim1.5$, the normalization factor in Eq.~(\ref{eq:mffPL}) is considerably smaller than what follows from integration of a log-normal PDF. 
Only at the highest densities or at very low Mach numbers do LNs tend to contain less mass than PLs.  As a result, for most PLs in most extragalactic regions, the mass predicted to form stars above the critical density is reduced, ultimately yielding efficiencies back down at the 1\% level, even while assuming $\epsilon_{\rm core}=1$ \citep[see also][\citetalias{burkmocz}]{burk18}. 

The multi-free-fall normalization factor in Eq.~(\ref{eq:mffPL}) is also independent of Mach number, which now functions solely to set the critical density for core formation in this scenario.  MFF PL predictions for $\epsilon_{\rm ff}$ thus share the much weaker reverse dependence on Mach number characteristic of the original \citetalias{KM05} models (see Figure \ref{fig:PHANGSsfe}), rather than the strong increase implied in Eq.~(\ref{eq:mffLN}).  These models instead achieve a large dynamic range in $\rm \epsilon_{\rm ff}$ through a strong dependence on power-law index $\alpha$ (or $k$) \citep{parmentier19,parmentier20, burk18, meidt20}.  
Resolved local clouds do exhibit a link between $\rm \epsilon_{\rm ff}$ and $\alpha$ (increasing $k$) like that predicted here \citep{burk18}.  In the remainder of this work, in the context of the galactic bottleneck we interpret variations in extragalactic efficiencies mainly as a result of variations in internal power-law structure.

It is worth emphasizing that, like pure PLs, hybrid non-smooth PDFs with $s_t\sim 1-3$ and a PL slope in the observed range $\alpha\sim 1.5-2.5$ exhibit a deficit of intermediate-density material and a downward kink appearance characteristically in regions with the relatively high Mach numbers ($\mathcal{M}\sim 8-100$) common to cloud-scale observations of extragalactic gas.  This presents an interesting contrast to hybrid smooth PDFs in this regime, especially in light of the way gravity is conventionally viewed, i.e. as building a PL on top of a LN, creating structure that resembles a hybrid smooth PDF or a PDF that at most kinks upward \citep{girichidis14,khullar22,jaupart}.  At high Mach numbers, such hybrid smooth-PDFs with no kink and $s_t\sim 1-3$ %(as in the hybrid smooth-PDF) 
would let gravity amplify the density contrasts seeded by turbulence in way that the PL slope is significantly shallower than $\alpha=1.5$.\footnote{Conversely, at high Mach number ($\mathcal{M}\sim 8-100$), hybrid smooth-PDFs with $\alpha\sim 1.5-2.5$ but no kink would need to have $s_t\sim 4-20$, which lies well beyond typical estimates for $s_{crit}$ for the highest Mach numbers.}  We regard such PDFs as not necessarily more physical than downward kink PDFs, 
not least because, unlike the way it can be simulated, the gas is never in a `non-self-gravitating' initial state with a LN PDF that it evolves away from.  Self-gravity shapes the gas distribution from the start, immediately shifting any intermediate-density material built up by turbulence to higher densities, shaping the PDF into a PL with $\alpha\sim 1.5-2.5$.  We thus favor the added flexibility of the hybrid non-smooth PDF over the smooth PDF in what follows, although we test both.  As discussed later in $\S$~\ref{sec:results}, expanded measurements of dense gas fractions will offer powerful insight into the internal structure of gas especially at the high Mach numbers typical in clouds outside of the Solar Neighborhood. 

\subsubsection{A finite reservoir to undergo collapse and replenish the PDF}\label{sec:tstopmodel}
A further characteristic of the secondary galactic bottleneck is its consequence for the renewal of gas density structure over time.   Following the arguments in $\S$ \ref{sec:trenew}, we assume that the density PDF is in a steady-state only until the free-fall time at the threshold density for self-gravitation $\rho_{\rm G}$, i.e. $t_{\rm stop}=t_{\rm ff,G}$.  For times $t>t_{\rm stop}$, star formation effectively shuts off, since the remaining gas below $\rho_{\rm G}$ is not strongly self-gravitating and unavailable for collapse (until perhaps the cloud environment is modified, introducing a new $t_{\rm stop}$).  These and other timescales used in this work are summarized in Table \ref{tab:timescales}.

Equating $t_{\rm stop}$ with the free-fall time at the threshold $\rho_{\rm G}$ given by Eq. (\ref{eq:collthresh}) chosen for study in this work, \footnote{\cite{meidt20} found that the rate of collapse $t_{\rm G}$ at $\rho_{\rm G}$ is slightly longer than the free-fall time %(approaching $t_{\rm ff}$ with increasing density), 
i.e. $t_{\rm G}\sim 2.4 t_{\rm ff}$. } 
 we let 
\begin{equation}
\frac{t_{\rm stop}}{t_{\rm ff}(\rho_0)}=\frac{\gamma_0}{\gamma_{\rm G}} \label{eq:tstop}
\end{equation}
or generically, in terms of the virial parameter $\alpha_{\rm vir}$, 
\begin{equation}
\frac{t_{\rm stop}}{t_{\rm ff}(\rho_0)}=\frac{1}{\gamma_{\rm G}}\frac{1}{(\alpha_{\rm vir}-1)^{1/2}}. \label{eq:tstopalpha}
\end{equation}

With these estimates for $t_{\rm stop}$, predictions for $\rm \epsilon_{\rm ff}$ from all varieties of turbulence-regulated star formation models that take into account the dynamical coupling of gas to the host galaxy potential are thus modified to read
\begin{eqnarray}
\epsilon_{\rm ff}&=&\epsilon_{\rm ff,steady}\frac{\gamma_0}{\gamma_{\rm G}}\frac{t_{\rm ff}(\rho_0)}{t_{\rm obs}} \textrm{\hspace*{0.5cm}for $\gamma_0<\gamma_{\rm G}$}\nonumber\\
&=&\epsilon_{\rm ff,steady} \textrm{\hspace*{2.25cm}for $\gamma_0\gg\gamma_{\rm G}$  } \label{eq:SFEstop}
\end{eqnarray}
in cases where $t_{\rm stop}$$<$$t_{\rm obs}$$\lesssim$$t_{\rm ff}(\rho_0)$.   
Here $\epsilon_{\rm ff,steady}$ is the value of the efficiency predicted under the assumption that the entire cloud will undergo collapse in a cloud free-fall time $t_{\rm ff}(\rho_0)$ (i.e. the predictions given in the previous sections in the single or multi-free-fall scenarios with LN or PL PDFs). 

Note that this impact of the secondary bottleneck on the $\epsilon_{\rm ff}$ is only expected to be detectable with long-timescale star formation tracers, $t_{\rm obs}>t_{\rm stop}$.  Galactic clouds likely fall in the second regime above, either because of the short visibility timescales of typical star formation tracers (for which $t_{\rm obs}<t_{\rm stop}$) 
or when examined (preferentially) at high densities where the cloud material is already decoupled from motion in the galactic potential ($\gamma_0\gg\gamma_{\rm G}$).  Cloud populations probed with longer-visibility extragalactic star formation tracers, on the other hand, will tend to fall in the first regime.  In such cases, $t_{\rm obs}\approx t_{\rm ff}(\rho_0)$ (see $\S$ \ref{sec:tobs}), implying that the star formation rates for clouds that contain some amount of material that remains coupled to the galaxy can be measurably lower than prescribed by, e.g. Eqs. (\ref{eq:mffLN}) or (\ref{eq:mffPL}). 

The combination of Eq. (\ref{eq:SFEstop}) with Eq. (\ref{eq:mffPL}) used for $\rm \epsilon_{\rm ff,steady}$ is equivalent to the %from the 
prediction presented in \cite{meidt20}, under the assumption that $t_{\rm obs}\sim t_{\rm ff}$.  In that work, the fraction of the cloud mass available to star formation was  expressed in terms of the self-gravitating (or decoupled) gas mass fraction rather than the ratio of timescales introduced here. The predictions in the next section represent an improvement over the \cite{meidt20} predictions in several regards. 
First, whereas previously a fixed core formation efficiency was adopted, here we assign a value that depends on $s_{\rm crit}$ estimated from observed gas properties.  Second, in this work we leverage variations in gas density structure to improve the match to observations, rather than adopting a fixed $k$=2 as in \cite{meidt20}. 

\section{Testing modified turbulence-regulated star formation models with PHANGS}\label{sec:obsTests}
\subsection{Summary of the proposed modelling strategy}\label{sec:strategy}
In this paper we propose adjustments to turbulence-regulated models of star formation to improve their consistency with observations.  
Building from the turbulence-regulated threshold for core formation $\rho_{\rm crit}$ as predicted by \citetalias{KM05} and \citetalias{PN11}, the main features we propose implementing to match the observed range in $\epsilon_{\rm ff}$ and its variation with cloud-scale gas properties in practice are: 
\begin{itemize}
\item a broad power-law density PDF component to capture the onset of self-gravitation (motivated by observed cloud structure), in practice shifting the overall normalization of the multi-free-fall $\epsilon_{\rm ff}$ down and reproducing a weak inverse dependence on Mach number as observed and reproduced by the original SFF \citetalias{KM05} theory.   
\item multi-free-fall core formation (or PDF renewal) allowing for greater dynamic range than predicted in single-free-fall scenarios (\citetalias{FK12}), spanning from cores to clouds to starbursts  
\item finite PDF replenishment that limits the duration for star formation to less than a cloud free-fall time, contributing a further reduction in $\epsilon_{\rm ff}$ that can vary locally.  The model of renewal we consider in this work is the result of the galactic bottleneck hypothesized by \cite{meidt20}, which prevents a fraction of cloud material from collapsing to participate in the star formation process.  
\end{itemize}

\begin{figure*}[t]
%\begin{flushleft}
%\vspace*{-.15in}
\begin{center}
\begin{tabular}{cc}
\hspace*{-.7in}\includegraphics[width=0.65\linewidth]{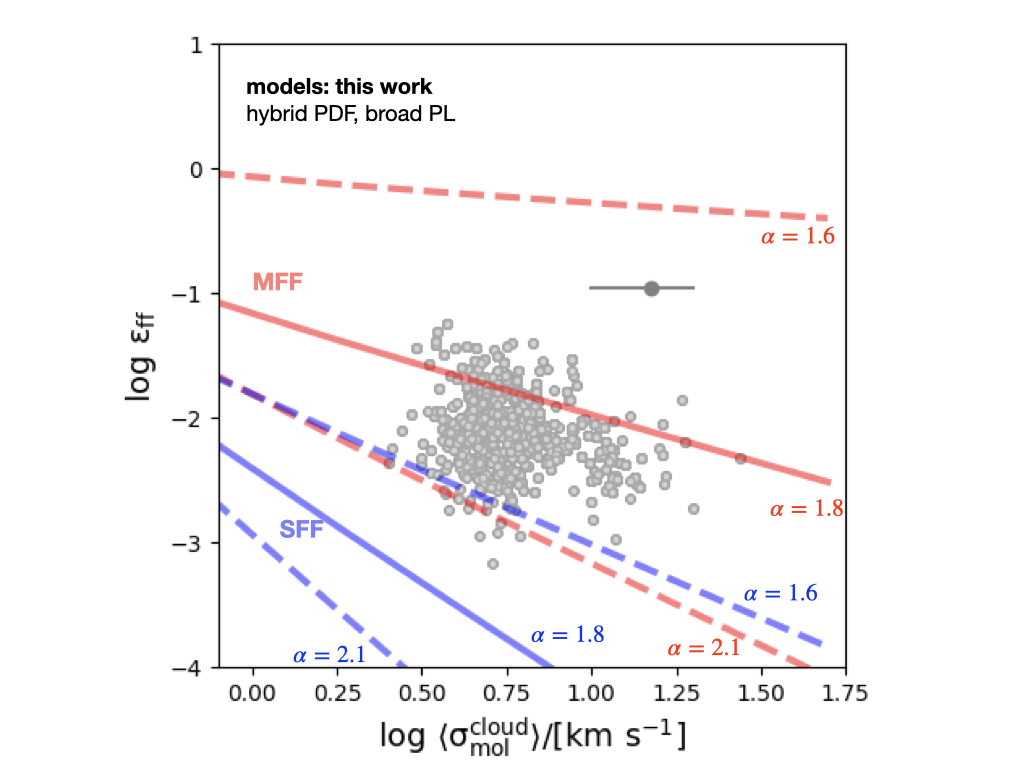}&%Bplot150nh.jpg}%modBplot150nh.jpg}%pdf}
\hspace*{-0.9in}\includegraphics[width=0.65\linewidth]{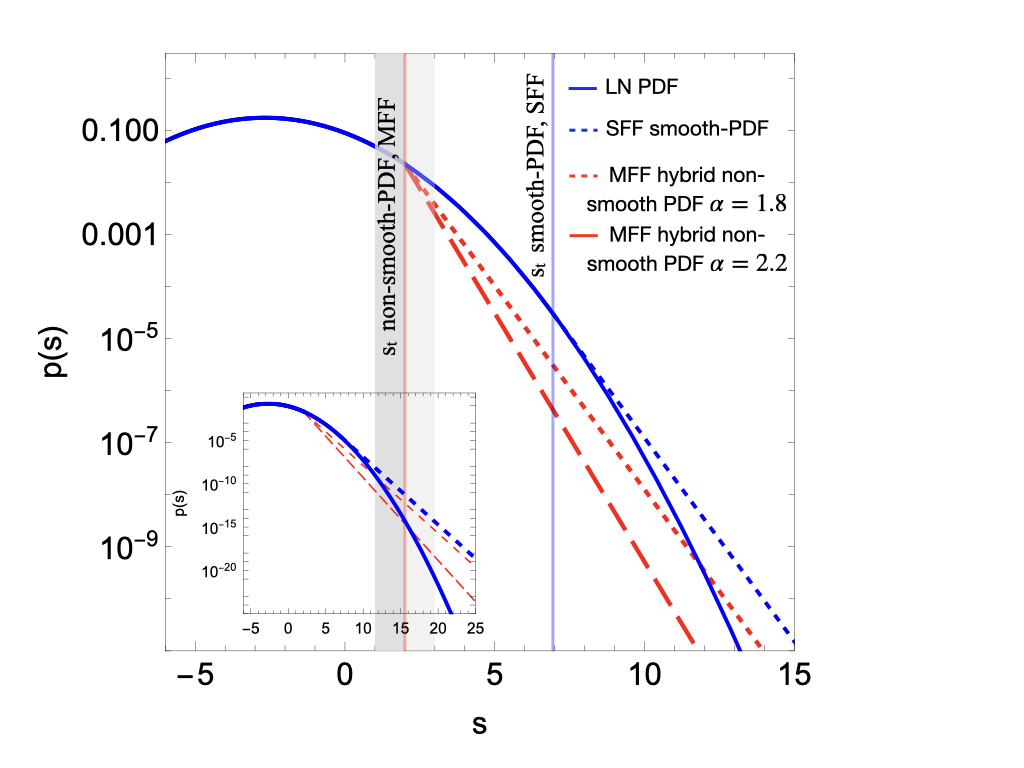}%Bplot150nh.jpg}%modBplot150nh.jpg}%pdf}
\end{tabular}
\end{center}
%\vspace*{-.15in}
\caption{(Left) Predictions for $\rm \epsilon_{\rm ff}$ from turbulence-regulated SF models with the hybrid LN+PL PDF proposed here (Eq. \ref{eq:finaleff}) in the single-free-fall (blue) or multi-free-fall (red) scenarios.  In these hybrid PDFs, the transition from lognormal to power-law behavior is set to density threshold for gas to kinematically decouple from the galaxy \citep{meidt20}.  A range of power-law slopes $1.6<\alpha<2.1$ set to the range observed by \cite{Kainulainen14} and \cite{schneider22} are indicated by the width of each band. As in Figure \ref{fig:Burksfe}, here we set $\epsilon_{\rm core} =0.5$, $\phi_t=1$, $\alpha_{\rm vir}=5$, $b=0.87$ and $c_s = 0.3 {\rm km \, s^{-1}}$ and use the \citetalias{KM05} critical density in Eq. (\ref{eq:scritKM05})  Here we also adopt a fixed $\gamma=1$.   
Light gray points show the PHANGS measurements from \cite{leroy24} and the dark gray bar and point depicts the \cite{dz23} $z\sim1$ clumps, repeated from Figure \ref{fig:PHANGSsfe}.   (Right) Illustration of typical hybrid LN+ PL density PDFs that can fit the observed $\rm \epsilon_{\rm ff}$.  All cases adopt the average cloud-scale velocity dispersion $\langle\sigma_{\rm mol}^{\rm cloud}\rangle=$~5 km s$^{-1}$, corresponding to $\mathcal{M}=16.7$. The two red lines show the non-smooth hybrid PDFs that fit in a multi-free-fall scenario. (The non-smooth PDF required in single-free-fall scenario is not shown.) These assume the same transition density $\rho_t=\rho_{\rm G}$ corresponding to $s_t=2$ (red vertical line) but the dotted PL has a slope $\alpha=1.7$ and the dashed PL has a slope $\alpha=2.2$, selected to bracket the full range in $\epsilon_{\rm ff}$ covered by PHANGS (see left panel).  The two blue lines represent either the SFF smooth hybrid PDF from Figure \ref{fig:Burksfe} (blue with dashing) or a LN-only PDF (solid blue).  The vertical blue line marks the transition $s_t=6.9$ for the former.  The two gray vertical lines are as plotted in Figure \ref{fig:Burksfe}.  The inset at the bottom left shows all four PDFs over a larger dynamical range, highlighting the behavior at the very highest densities.  
  }
\label{fig:newmodelsfe}
%\end{flushleft}
\end{figure*}
The second and third of these determine the behavior of the free-fall factor during integration over the density PDF and affect the normalization of the predicted $\epsilon_{\rm ff}$. The first involves a hybrid PDF identical in form to the model proposed by \citetalias{burkmocz} (see Eq.~[\ref{eq:hybridPDF}]), with identical normalization.  % as determined by Burkhart.  
In what follows, though, we omit the link between the power-law index and $s_t$ required by the differentiability condition adopted by \cite{burk18}.  Instead, we test the $s_t$ value implied by the criterion for self-gravity to dominate the external galactic potential determined by \cite{meidt20}, i.e. $s_t=s_{\rm G}=\ln (\rho_G/\rho_0)$ with $\rho_G$ given by Eq. (\ref{eq:collthresh}).   This leaves $\alpha$ as a free parameter.  

In general, and as argued in $\S$ \ref{sec:powerlawchange}, in the disks of star-forming galaxies we expect $s_{\rm G}<s_{\rm crit}$,  with $s_{\rm crit}$ related to the cloud-scale turbulent gas properties (\citetalias{KM05}).  We thus restrict star formation to gas in the PL tail.  
The $\epsilon_{\rm ff}$ in this case can be determined by comparing the mass in stars formed, calculated by integrating above $\rho_{\rm crit}$, to the total gas mass, calculated by integrating the PDF above $\rho_c$\footnote{Following convention, the hybrid PDF is normalized so that $\int_{-\infty}^{\infty} p(s)ds=1$.  But unlike in the case of a pure log-normal PDF that is assumed to be centered at $s_0=-\sigma_s^2/2$, the mass $\int_{-\infty}^{\infty} e^s p(s)ds$ in the case of the hybrid, normalized PDF (either smooth or non-smooth) is not unity.}.  We find
\begin{equation}
\epsilon_{\rm ff}=\epsilon_{\rm core}\frac{\gamma_0}{\gamma_{\rm G}}%\left[
\frac{C f(\rho_{\rm crit},\alpha)}{\frac{1}{2}\left(1+\textrm{erf}\left[\dfrac{(s_t-\sigma_s^2/2)}{\sqrt{2\sigma_s^2}}\right]\right)+C\dfrac{e^{(1-\alpha) s_t}}{\alpha-1}}
%\right]
\label{eq:finaleff}
\end{equation}
with
\begin{equation}
C=\frac{1}{\sqrt{2\pi}\sigma_s}e^{-\dfrac{(s_t+\frac{1}{2}\sigma_s^2)^2}{2\sigma_s^2}+s_t\alpha}
\end{equation}
and
\begin{equation}
f(\rho_{\rm crit},\alpha)=
\begin{cases}
 \dfrac{e^{s_{\rm crit}(\frac{3}{2}-\alpha)}}{\alpha-3/2}& \text{MFF}\\
\dfrac{e^{s_{\rm crit}(1-\alpha)}}{\alpha-1}& \text{SFF}.
\end{cases}
\end{equation}
This assumes that the LN component of the PDF has width $\sigma_s$ given by Eq. (\ref{eq:sigmat}).  For compatibility with the \cite{burk18} formulation framed around the smooth hybrid LN+PL PDF in Eq.(\ref{eq:hybridPDF}), here $\phi_t$ is omitted (set to unity).  

The factor $\gamma_0/\gamma_G$ in eq.~(\ref{eq:finaleff}) reflects the assumption that PDF renewal lasts only until $t_{\rm stop}=t_{\rm ff,G}<t_{\rm ff,0}$, where $t_{\rm ff,G}$ is the free-fall time at the density $\rho_{\rm G}$ where self-gravity dominates the evolution and we have set the observed duration of the clouds to their lifetime, which we further assume to be near the free-fall time at the mean cloud density $t_{\rm ff,0}$.  When the duration of PDF renewal lasts for the full cloud lifetime, $t_{\rm stop}=t_{\rm ff,0}$ and the factor $\gamma_0/\gamma_G$ is omitted, as discussed in $\S$~\ref{sec:tstopmodel}. In the model we choose to test in what follows, $\rho_{\rm G}$  is given by threshold where gas is decoupled from the galactic potential (see $\S$ \ref{sec:M20bottleneck}).  We also use this density to define $s_t$ that marks where the density PDF transitions from LN to PL.  

As result of these changes motivated by the introduction of self-gravity, the output efficiency predicted by MFF scenarios is greatly reduced.  This is illustrated in Fig.~\ref{fig:newmodelsfe} showing predictions from Eq. (\ref{eq:finaleff}) with $\epsilon_{\rm core}=0.5$ in both the MFF and SFF scenarios where (for illustration purposes) a universal threshold $\rho_{\rm G}$ is adopted and the power-law slope is fixed to $1.5<\alpha<2.5$ (as in Figure \ref{fig:Burksfe}).  Compared to either Figure \ref{fig:PHANGSsfe} or Figure \ref{fig:Burksfe} the MFF predictions provide the closer match to the observed $\rm \epsilon_{\rm ff}$, all with realistic density PDFs.  We call these predictions self-gravitating MFF predictions, or sgMFF predictions, in what follows.  
In the SFF scenario, on the other hand, exchanging a LN PDF for a broad PL tail leads to too much reduction in $\epsilon_{\rm ff}$, and the inclusion of the factor $t_{\rm stop}/t_{\rm obs}$ describing the galactic bottleneck increases the discrepancy with the observations further.  These predictions using a hybrid PDF with a broad PL tail are referred to as sgSFF predictions.  

To obtain the best match to the observations, it is clear that the power-law slope $\alpha$ must vary systematically throughout the targets, i.e. as a function of $\sigma$ or related property; no single value of $\alpha$ covers the observed range in $\rm \epsilon_{\rm ff}$.  
Although there are models, e.g. for the evolution of $\alpha$ over time \citep[e.g.][]{murray, lee15, caldwell18}, at this point we prefer not to translate expectations into a dependence on cloud-scale properties and instead opt to fit Eq.~(\ref{eq:finaleff}) to the data to determine the $\alpha$ that is necessary to match the observations everywhere.  By comparing the fitted behavior to independent expectations, we can then qualitatively assess the model ($\S$.  \ref{sec:discussion}).  

\subsection{Overview of empirical tests}\label{sec:empiricalTests}

The $\epsilon_{\rm ff}$ model given in Eq.~(\ref{eq:finaleff}) contains a number of factors that can all either be constrained directly with observations of molecular gas on cloud scales or have been shown (observationally or with simulations) to fall within relatively narrow ranges (i.e. $b$, $\epsilon_{\rm core}$ and $\alpha$).  
For the remainder of this section, we will describe how we use cloud-scale information constrained by PHANGS (compiled by \citealt{sun22,sun23} and summarized in $\S\S$ \ref{sec:phangssfes} and \ref{sec:PHANGS}) to model the time-averaged $\epsilon_{\rm ff}^{\rm obs}$ measured as in Eq. (\ref{eq:obsSFE}) by \cite{leroy24} within kpc-size regions sampling throughout PHANGS targets using a long-visibility ($\sim$10-100 Myr) star formation tracer.  For a discussion of the impact of the CO-to-H$_2$ conversion factor on measured cloud properties and the estimated $\epsilon_{\rm ff}$, as well as the role of the completeness correction recommended by \cite{leroy24}, see $\S$ \ref{sec:phangssfes}. 

\subsubsection{Using the comparison between the model and the observations to constrain $\alpha$}
For comparing to PHANGS observations, we constrain $\sigma_s$, $s_{\rm crit}$ and $s_{t}=\ln{\rho_g/\rho_0}$ empirically (via eqs. [\ref{eq:sigmat}], [\ref{eq:scritKM05}], [\ref{eq:collthresh}], respectively, with the average cloud-scale velocity dispersion and gas volume density measurements as inputs), and leave the power-law index as the only remaining variable.  With our approach -- solving for the value of $\alpha$ that best matches the predictions to the observations -- our determinations of $\alpha$ are only as good as the $\epsilon_{\rm ff}$ model, and they inherit the uncertainties associated with factors that are either not well-constrained or incorporated into the model at present.  
For the extragalactic measurements we consider in this work, the value of $t_{\rm obs}$, for one, requires additional arguments that may make it subject to greater uncertainty than in the case of shorter-timescale galactic star formation tracers (see $\S$ \ref{sec:tobspresc} below).  As part of our assessment of the quality of the model in Eq.~(\ref{eq:finaleff}), we thus also examine the $\alpha$ values implied by three alternative $\epsilon_{\rm ff}$ models.  

\subsubsection{The optimized hybrid LN+PL PDF $\epsilon_{\rm ff}$ models}
In addition to testing the $\epsilon_{\rm ff}$ model given by Eq.~(\ref{eq:finaleff}) as both sgMFF and sgSFF, we test the scenario in which the duration of star formation is not restricted to below the cloud free-fall time and assume $t_{\rm stop}\approx t_{\rm ff}$.  
We also compare with MFF and SFF predictions using the hybrid smooth-PDF of \cite{burk18} and \cite{burkmocz} with $s_t=s_{\rm crit}$, adopting identical values for all mutual input parameters.  In all four tested models, we adopt $b=0.87$ for the turbulent driving parameter, $c_s=0.3$~km s$^{-1}$ and $\epsilon_{\rm core}=0.5$ but comment on the impact of variation in these parameters, where applicable.  

The value of $\alpha$ that best matches each of the four models to an observed $\epsilon_{\rm ff}^{\rm obs}$ is determined using non-linear least squares regression, specifically the {\tt curve\_fit} routine in the {\tt scipy.optimize} \python~package.  The regression is bounded to $1.5<\alpha<5$ and given an initial input of 
$\alpha=1.8$, the mean value determined for the resolved clouds analyzed by \cite{Kainulainen14}.  Note that width of the LN portion of each fitted model is constrained by the observations, but its normalization and high density extent depends on the fitted $\alpha$ (see Eq.~\ref{eq:finaleff}).

With this optimization, estimates of $\alpha$ are determined by fitting to groups/subsets of kpc-scale $\epsilon_{\rm ff}^{\rm obs}$ measurements. 
To retain as much leverage on systematic variation in $\alpha$ as possible, we adopt a relatively fine sorting, first by galactocentric radius and then by the mass--weighted average cloud-scale molecular gas surface density $\langle\Sigma_{\rm mol}^{\rm cloud}\rangle$ measured in the associated kpc-sized region, motivated by the large dynamic range in $\langle\Sigma_{\rm mol}^{\rm cloud}\rangle$ present at a given radius \citep{meidt21,sun22,leroy24}. 

Specifically, after the measurements are sorted into a series of bins of galactocentric radius, each bin is further divided into three groups: below the 30$^{th}$ percentile of the molecular gas surface density distribution at that radius, between the 30$^{th}$ and 70$^{th}$ percentiles and above the 70$^{th}$ percentile.  This division yields groups with roughly equal numbers of (3 or more) measurements per group.  In subsequent calculations, all individual measurements in a given group are assigned the same value of $\alpha$ determined for that group.  
In testing, we have found that allowing for systematic variation in $\alpha$ of this kind yields a better match to the observations than when, e.g., all measurements at a given radius are used to constrain $\alpha$.  It is worth emphasizing that we have not exhaustively tested alternative sorting schemes (e.g. by velocity dispersion, stellar surface density, etc.), and there may be preferable approaches to the one we adopt here.  Our main goal is to provide a basic indication of variation in $\alpha$, where present, that can be examined in future work.

\subsubsection{Prescriptions for $t_{\rm obs}$\label{sec:tobspresc} appropriate for long visibility star formation tracers}\label{sec:tobs}
To compare our full $\epsilon_{\rm ff}$ sgMFF (or sgSFF) model in Eq.~(\ref{eq:finaleff}) to extragalactic measurements we must assign a representative $t_{\rm obs}$ appropriate for the long visibility timescales of extragalactic kpc-scale star formation tracers.  (The scenario we describe below does not apply when $t_{\rm sf}$ probes within the lifetimes of the individual clouds.) 
The star formation rates constructed on kpc-scales by PHANGS are designed to capture massive star formation over a range of phases and timescales.  Each hexagonal kpc region samples multiple instances of the star formation cycle and thus effectively captures star formation over the last $t_{\rm sf}\sim~100$ Myr \citep{leroy24}.  However, in this scenario, when the long the visibility timescale $t_{\rm sf}$ exceeds the time to complete any single star formation cycle, or roughly the lifetime of a typical GMC \citep{Chevance20}, then $t_{\rm obs}\neq t_{\rm sf}$.  Instead, at maximum $t_{\rm obs} = t_{\rm life} + t_{\rm no~cl}$, where $t_{\rm life}$ is the representative cloud lifetime (assumed to vary negligibly locally) and $t_{\rm no~cl}$ is the time between cloud formation events.\footnote{Note that this procedure is mathematically equivalent to setting $t_{\rm obs} = t_{\rm sf}$ while reducing the estimated SFR by a factor of $1 / (t_{\rm life} + t_{\rm no~cl})$ to account for the fact that our SFR tracer is sampling multiple generations of star formation.} That is, during the total time $t_{\rm obs}$ over which star formation events are being `observed', $t_{\rm no~cl}$ is the time that gas is not participating in the star formation process.  This factor %$t_{\rm no~cl}$ 
 allows us to properly take into account that, whereas 
our probe of the molecular gas surface density is only telling us about the current generation of clouds, our SFR tracer is also sensitive to the contributions of previous generations of star formation within the same region within the last $\sim 100$~Myr.  
 %To account for this, it is necessary to limit  
Note, though, that $t_{\rm no~cl}$ is distinct from the time clouds spend in an inert phase ($t_{\rm no~sf}$), which is contained within $t_{\rm life}$ and is responsible, i.e., for offsetting $t_{\rm start}$ from the moment of cloud formation or marking the cessation of star formation before a cloud free-fall time has elapsed.  We direct the reader to \cite{Kruijssen19}, \cite{Chevance20, Chevance22}, and \cite{kim22} for a description of how different phases of the cloud and star formation cycle can be measured observationally to place constraints on the responsible physical mechanisms.

In practice, we assume for the purposes of the discussion in this paper that the number of clouds within each (high-completeness) 1.5~kpc region is approximately conserved (i.e.\ that new clouds form at roughly the same rate as old clouds are destroyed), which implies that $t_{\rm no~cl}$ is negligible in comparison to $t_{\rm life}$.
Observational measurements of $t_{\rm life}$ place it within a factor of 2 of $t_{\rm ff}$ \citep{Chevance20}, suggesting that, with long-visibility tracers, the minimum window of time that each individual cloud is probed can be well approximated by the cloud free-fall time $t_{\rm ff}(\rho_0)$.  This should be good to within a factor of roughly two, but scenarios in which $t_{\rm no~cl}$ represents a non-negligible part of the cloud life-cycle may underestimate $t_{\rm obs}$ more strongly.  

\subsection{Additional empirical constraints from PHANGS}\label{sec:PHANGS}
To determine the critical threshold for self-gravitation, $\rho_{\rm G}$, for each `representative cloud' in a given measurement region, we develop an empirical estimate for the parameter $\gamma$ in Eq. (\ref{eq:gamma1}) that compares gas self-gravity to the strength of the background galaxy potential on the outer cloud scale $R_{\rm c}$.   

The self-gravitational potential energy in the numerator of Eq. (\ref{eq:gamma1}) is written in terms of the gas surface density $\langle\Sigma_{\rm mol}^{\rm cloud}\rangle$ on the cloud scale $R_{\rm c}$ assuming that the gas is arranged with a triaxial geometry (with cloud vertical extent $Z_c=qR_{\rm c}$ where the axis ratio $q\sim1$) and follows a power-law density distribution $\rho\propto r^{-k}$.  %Here the 

As derived in \cite{meidt18,meidt20}, on the cloud scale, the background (rotating) potential is written in terms of the effective pressure set up by the gas kinematic response to the associated gravitational forcing.  Appendix \ref{sec:appendixgamma} describes how we estimate the background potential, using empirical estimates of the underlying stellar density and observed gas rotational velocities.  

\begin{figure}[t]
%\begin{flushleft}
%\vspace*{-.15in}
\begin{center}
\begin{tabular}{cc}
\includegraphics[width=0.95\linewidth]{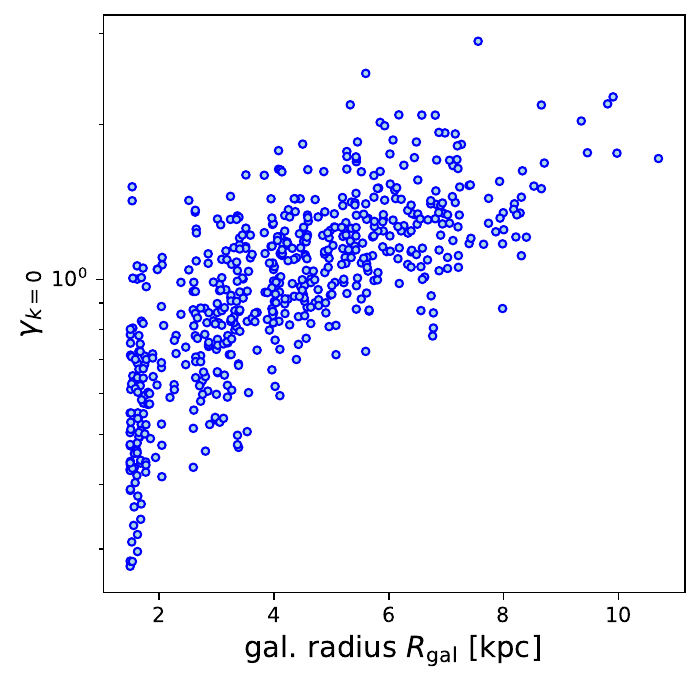}\\
\end{tabular}
\end{center}
%\vspace*{-.15in}
\caption{Cloud-scale values of $\gamma_{k=0}$ estimated with Eq.~(\ref{eq:gamma_0}) shown as a function of galactocentric radius for the subset of hexagonal apertures that sample within galaxies with fitted rotation curves (see Appendix \ref{sec:appendixgamma}). Moving inwards towards galaxy centers, the influence of the self-gravity of clouds weakens with respect to the galactic potential.}
\label{fig:gammas}
%\end{flushleft}
\end{figure}

Figure \ref{fig:gammas} shows the radial distribution of $\gamma$ measured on the 150-pc cloud scale throughout the target sample, adopting $k=0$.  Values range from $\gamma_{k=0}\sim0.5$ at inner radii to $\gamma_{k=0}\sim2$. From the median (mean) $\gamma_{k=0}=1.1$ ($\gamma_{k=0}=1.08$), we infer that gas self-gravity and the galactic potential are comparably strong on the cloud scale.   This has been suggested earlier using semi-empirical galaxy models \citep{meidt18,meidt20}.  As in those models, observed galaxy centers stand out as regions where the galactic potential becomes increasingly important relative to gas self-gravity.  These are thus regions where the effective pressure in the gas is dominated by an external component, despite the typically elevated gas surface densities in these zones.  
As a consequence, the gas density PDF would transition from LN to PL at a higher threshold than in the surrounding disk, signifying that less gas is available to participate in the star formation process over the course of a cloud free-fall time.  

\subsection{Results}
\label{sec:results}
\begin{figure}[t]
%\begin{flushleft}
%\vspace*{-.15in}
\begin{center}
\begin{tabular}{c}
\includegraphics[width=0.99\linewidth]{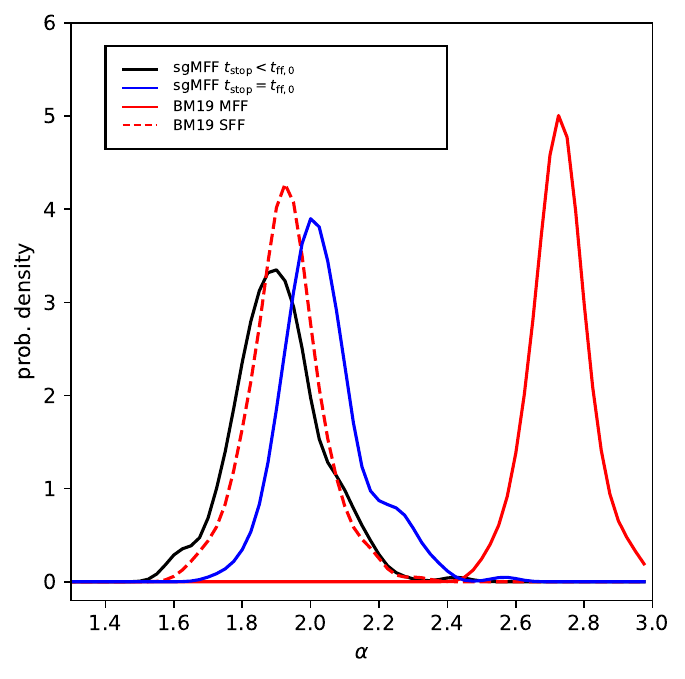}%jpg}%\\
\end{tabular}
\end{center}
%\vspace*{-.15in}
\caption{%(Top) 
Histograms of the power-law slope $\alpha$ in the hybrid (LN+PL) PDFs that match four different SF models (black, blue and red and red dashed) to the $\rm \epsilon_{\rm ff}$ measured in PHANGS.  The fiducial shortened duration broad PL MFF model is shown in black, the full duration broad PL MFF model is shown in blue, the Burkhart smooth-PDF MFF model is shown in red and the Burkhart smooth-PDF SFF model is shown as red with dashing.%(Bottom) 
 }
\label{fig:pdfs}
%\end{flushleft}
\end{figure}

Figure \ref{fig:pdfs} summarizes the density structure required to fit PHANGS measurements of $\epsilon_{\rm ff}^{\rm obs}$ with four different SF models, each pivoting on a hybrid LN+PL density PDF and adopting $\epsilon_{\rm core}=0.5$.   In fitting, the width of the LN portion in all models is set by the local cloud-scale Mach number estimated from the observed gas velocity dispersion, assuming $c_s=0.3$~km s$^{-1}$ and $b=0.87$.  
In two out of the three MFF models (our fiducial models, with star formation duration either shortened $t_{\rm stop}<t_{\rm obs}$ or full $t_{\rm stop}\approx t_{\rm obs}$, portrayed in black and blue in Figure \ref{fig:pdfs}, respectively), the transition density $\rho_t$ between the LN and PL portions is the same and set to the density where the gas is decoupled from the external potential $\rho_{\rm G}=\rho_0(\gamma_{\rm G}/\gamma)^2$. 
In the third MFF model (portrayed by the red line in Fig. \ref{fig:pdfs}) as well as in the SFF model (portrayed by the red dashed line in Fig. \ref{fig:pdfs}), the transition density $\rho_t$ is tied to the slope of the power-law, as envisioned by \cite{burk18} (see Eq. [\ref{eq:burkst}]), and determined by the $\alpha$ selected during fitting.  

\subsubsection{Density structure}\label{sec:densstructure}
The $\alpha$ values in the four models are broadly similar, all falling comfortably in the range $\alpha=1.5-3$.  This would correspond to $k=1-2$ assuming the mass is arranged with a spherically symmetric radial density profile $\rho\propto r^{-k}$ where $k=3/\alpha$ \citep{girichidis14}.  Within this narrow range of $\alpha$ (or $k$), though, the four models entail appreciably different gas structure.  
In the MFF smooth-PDF model, which has the highest values for $\alpha\sim2.5-3$ ($k=1-1.2$), the transition from LN to PL is placed at densities $\ln (\rho_t/\rho_0)\sim10-15$, several orders of magnitude higher than in the fiducial model ($\ln (\rho_t/\rho_0)\sim1-3$;~see~$\S$~\ref{fig:gammas}) for the regions observed in PHANGS.  The PDFs in these smooth-PDF MFF models remain LN for longer (up to higher densities), requiring a dramatic reduction in the highest density content (through a much elevated $\alpha$) in order to match the observed $\epsilon_{\rm ff}$.  

In the fiducial model, in contrast, the more prominent power-law (and reduced high density content) makes it possible to match the $\epsilon_{\rm ff}$ with lower $\alpha\sim1.6-2.5$ (higher $k\sim1.2-1.9$) that are better consistent with the observed range \citep[$k$$\sim1.3-2$;][]{Kainulainen14}.  This is aided by the limitation to core formation in the shortened duration $t_{\rm stop}<t_{\rm ff,0}$ model, which allows $\alpha$ to reach the lowest values near $\alpha\sim$ 1.6 and still match the observed $\epsilon_{\rm ff}$.  

As illustrated in the right panel of Figure \ref{fig:newmodelsfe}, the form of the PDF and the role of the PL switches from high to low Mach number.  In the high Mach number regime where the majority of PHANGS observations sit, the PL kinks downward from the LN, bringing MFF models with hybrid PDFs into better agreement with extragalactic observations than pure PLs (see $\S$ \ref{sec:broadPL}).  In the low Mach number regime, the PL kinks upward, adding high density material relative to the pure LN.  The density PDFs for many local clouds \citep{schneider22} tend to show an upward kink, suggesting that in this regime, at low Mach number, the role of the PL is to enhance the predicted $\epsilon_{\rm ff}$ compared to an equal-mass LN PDF, unlike in extragalactic high Mach number clouds.

\begin{figure*}[hpt!]
%\begin{flushleft}
\hspace*{-.1in}
%\vspace*{-.4in}
% \begin{center}
% \begin{tabular}{c}
\centering
\includegraphics[width=.98\linewidth]{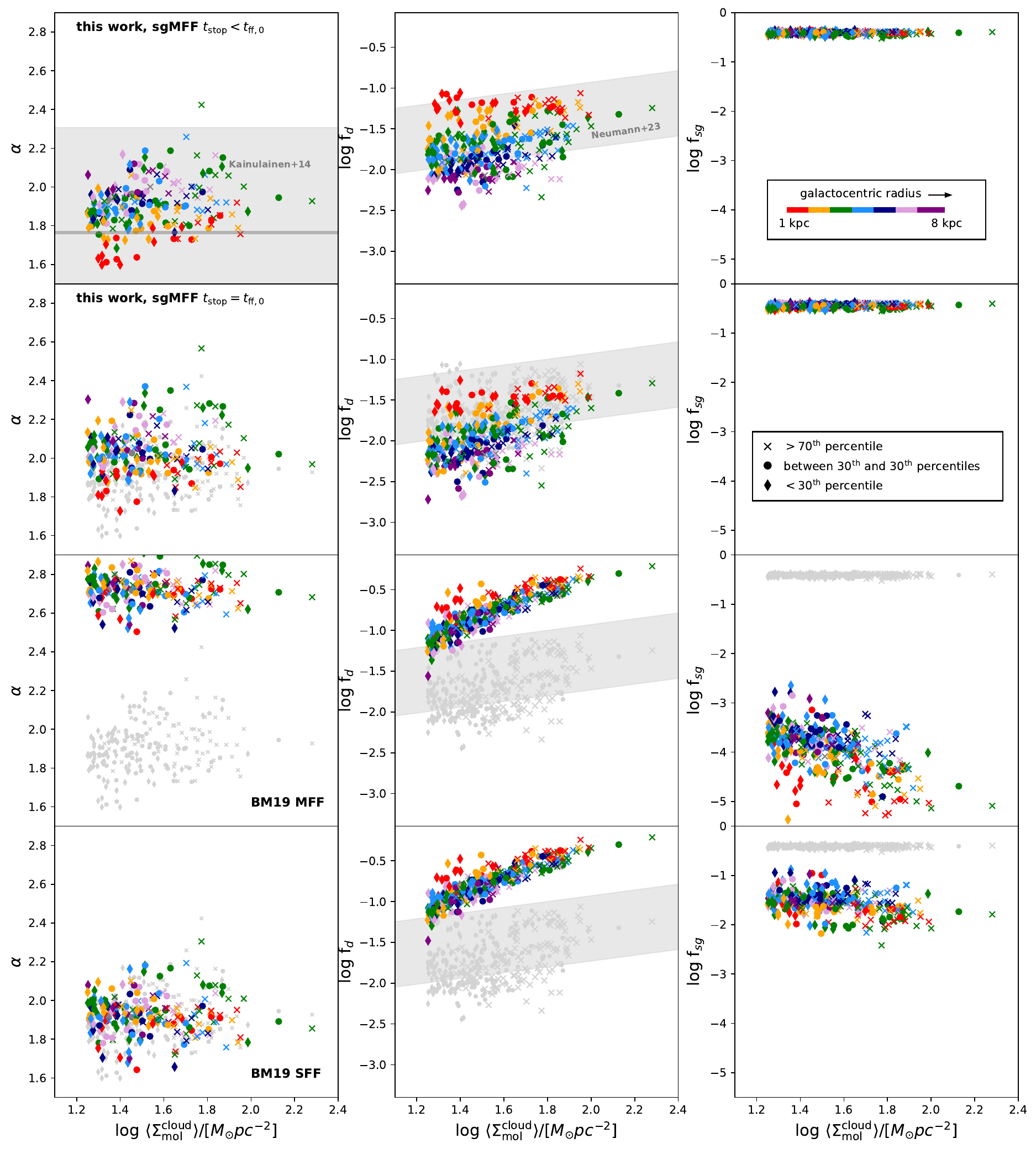}%150nh.jpg}
% \end{tabular}
% \end{center}
%\vspace*{-.1in}
\caption{Diagnostics of the hybrid LN+PL density PDFs that match different MFF (and SFF) SF models to the $\rm \epsilon_{\rm ff}$ values measured in PHANGS, from left to right: the slope of the PL component of the PDF $\alpha$, the dense gas fraction $f_{\rm d}$ measured above a fixed density threshold (see text) and the self-gravitating fraction $f_{\rm sg}$ measured above the critical density $s_{\rm crit}$. The top row shows results for PDFs in the fiducial broad PL, shortened duration model given in Eq.~(\ref{eq:finaleff}), adopting a MFF scenario. The second row shows results for PDFs in the broad PL, full duration model with $t_{\rm obs}=t_{\rm stop}$ in Eq.~(\ref{eq:finaleff}) also in the MFF scenario.  The bottom two rows show MFF (third row) or SFF (fourth row) predictions adopting hybrid smooth-PDFs in the Burkhart formulation, with $s_{\rm t}=s_{\rm crit}$ advocated by \cite{burkmocz}.  All SF models shown here have assumed $\epsilon_{\rm core}=0.5$, $\phi_t=1$, $b=0.87$ and $c_s=0.3$~km~s$^{-1}$ and the critical density given in Eq. (\ref{eq:scritKM05}).  
As indicated by the color bar in the top right panel, points are color-coded by galactocentric radius (increasing outward, from red to purple). Symbol style denotes relative gas surface density; in all radial bins, regions sitting below the 30$^{th}$ percentile are marked with diamonds, regions between the 30$^{th}$ and 70$^{th}$ are marked with a circle, and regions above the 70$^{th}$ percentile are marked with a cross.  The gray points in the bottom row repeat the measurements from the top row.  The horizontal gray line (band) in the top left panel shows the mean (full spread) of the PL slopes $\alpha$ observed in local clouds \citep{Kainulainen14}.  The gray band in the middle panels illustrates the relation between dense gas fraction and gas surface density fitted by \cite{neumann23}. The width of the band represents the full spread in the fitted data (roughly 3 times the scatter about the fit line).
\label{fig:results1}
}
%\end{flushleft}
\end{figure*}
\subsubsection{Dense gas content}\label{sec:denscontent}
The substantial differences in density structure implied by the four models are explored further in Fig.~\ref{fig:results1}, which show diagnostics of the density distribution in each model.  The bottom  row shows the diagnostics for the SFF smooth-PDF model for reference.  
In the left column, $\alpha$ is plotted against the cloud-scale gas surface density $\langle\Sigma_{\rm mol}^{\rm cloud}\rangle$.  
The middle column shows the dense gas fractions $f_d$ measured from each hybrid PDF above a fixed dense gas density threshold $n_d=10^3$ cm$^{-3}$, near the effective critical density of HCN(1-0) \citep{leroydense}.  The right column shows the self-gravitating gas fractions $f_{\rm sg}$ measuring the fraction of gas in the PL portion of the density PDF.  

Small differences in $\alpha$ amount to notable changes in the amount of the densest material.  The top middle panel exhibits good correspondence with the observed values of the dense gas fraction traced by HCN/CO and its empirical relation to gas surface density found by \cite{neumann23}.  
That trend resembles the linear relation expected for the fraction of gas above a fixed threshold in clouds with simple power-law distributions $\rho\propto r^{-k}$ with $k=2$ ($\alpha=1.5$), in particular.  The lower $\alpha$ values typical of the fiducial short duration model indeed help match the observed $f_d$ level better than the full duration models where $\alpha$ is slightly larger ($k$ is slightly smaller) on average.  

In the nominal models (top two rows of Figure \ref{fig:results1}), the PL represents a large portion of the total mass distribution (roughly a third to a half; see the rightmost column in Figure \ref{fig:results1}), such that the overall behavior resembles that of pure power-laws.    
In the hybrid smooth-PDF MFF and SFF models (bottom two rows of Figure \ref{fig:results1}), on the other hand, the PL portion is strongly limited.  Self-gravitating fractions are several orders of magnitude lower than in either of the two fiducial models (compare the top two rightmost panels with the bottom two rightmost.)  As a result, the dense gas fractions for the smooth-PDFs are driven by the LN component.  This not only produces a different (weaker) relation to $\langle\Sigma_{\rm mol}^{\rm cloud}\rangle$ than in the pure PL case, but tends to produce higher $f_d$ values, for the reasons discussed earlier in $\S$ \ref{sec:broadPL}, and less scatter as well.  So whereas an overall higher $\alpha$ would lower the dense gas fractions on average for pure PL PDFs, in the context of the smooth-PDF models, the overall higher values of $\alpha$ are associated with significantly higher dense gas fractions (i.e. in the third row in Fig. \ref{fig:results1}). 

A higher assumed density threshold could alleviate some of the mismatch between the $f_d$ estimated from the MFF and SFF smooth-PDF models and the values observed (and at the same time shift the fiducial models out of agreement with the observed range in $\alpha$).  But the absence of a prominent, shallow power-law would still make the MFF and SFF smooth-PDFs defined by \cite{burk18} less consistent with observed cloud density structure than reproduced with our fiducial models.  

In light of these results, we conclude that MFF scenarios provide their best match to observed cloud populations when the density distribution includes a prominent PL component, requiring no artificial reduction in $\epsilon_{\rm core}$.  The model for the transition density $\rho_t$ we examine here is able to satisfy this requirement, but there may be other mechanisms that act similarly.  
To get a sense for the validity of our proposed model, in the next section we will discuss the variations in $\alpha$ that it implies in relation to other expectations for gas density structure. 

\subsection{Discussion: systematic variation in $\alpha$}\label{sec:discussion}
\begin{figure}[t]
%\begin{flushleft}
%\vspace*{-.15in}
\begin{center}
\begin{tabular}{c}
\includegraphics[width=0.95\linewidth]{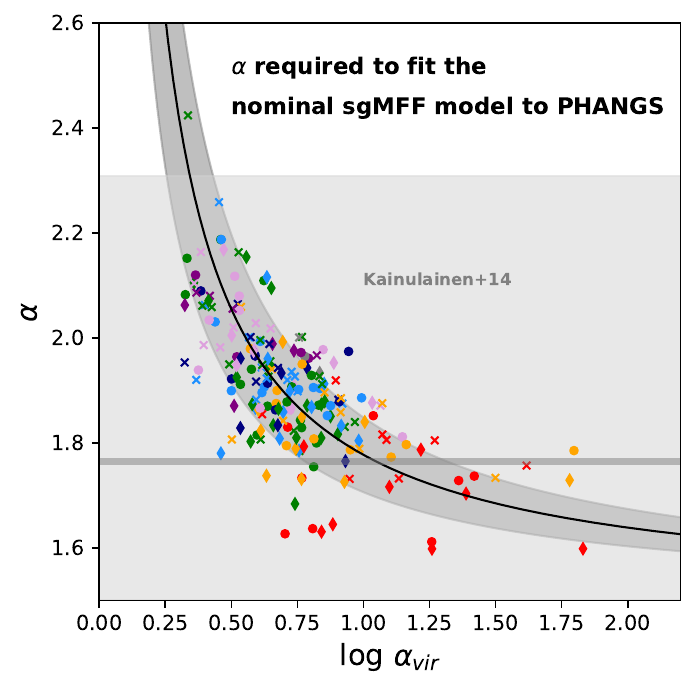}%jpg}%\\
\end{tabular}
\end{center}
%\vspace*{-.15in}
\caption{%(Top) 
 Variation in the PL slope $\alpha$ calculated with the shortened duration model given in Eq.~(\ref{eq:finaleff}) as a function of the cloud-scale virial parameter $\alpha_{\rm vir}$.  Symbol shapes and colors are as in Figure \ref{fig:results1}. % and  \ref{fig:results2}.  
 The horizontal gray line (band) shows the mean (full spread) of the PL slopes $\alpha$ observed in local clouds \citep{Kainulainen14}. The reference line given by Eq. (\ref{eq:approximatealpha}) with $\mathcal{R}_\alpha=0.23\pm0.06$ is shown in black. Combining the systematic variation of dense gas fraction with virial state implied here together with variations in the critical density (depending on $\alpha_{\rm vir}$) produces star formation efficiencies that match the PHANGS observations in Fig. \ref{fig:PHANGSsfe} (see also Fig. \ref{fig:PHANGSsfeMODEL} and the discussion in $\S$ 
 \ref{sec:discussionMFFvKM05}).
 }
\label{fig:alphaAlphavir}
%\end{flushleft}
\end{figure}
One of the trends revealed in Figure \ref{fig:results1} is the behavior of $\alpha$ at fixed radius (fixed color), which subtly but systematically decreases with decreasing $\langle\Sigma_{\rm mol}^{\rm cloud}\rangle$, moving from regions with locally higher than average surface density (crosses) to regions of locally lower than average surface density (diamonds). 
All three models show this to some degree and it is the result of the tight link between efficiency and power-law slope predicted by Eq. (\ref{eq:finaleff}) (and Eq. [\ref{eq:mffPL}]).  Clouds with lower than average surface density but forming stars at the same rate as clouds of higher than average density will appear to form stars more efficiently and, in our nominal LN+PL model, that higher efficiency is attributed to a higher content of dense gas.  The variation in $\alpha$ with $\langle\Sigma_{\rm mol}^{\rm cloud}\rangle$ is largest and clearest in the fiducial short duration model and weakens when the factor $t_{\rm stop}/t_{\rm obs}=\gamma/\gamma_{\rm G}$ is omitted.    

In the fiducial model, the decrease in $\alpha$ with decreasing $\langle\Sigma_{\rm mol}^{\rm cloud}\rangle$ occurs alongside a clear decrease in $\alpha$ moving towards smaller galactocentric radius (i.e. at fixed surface density percentile).  Both the virial level and gas pressure are observed to increase with decreasing galactocentric radius \citep{sun18,sun22}.  The implication of power-law models is therefore that high pressure, super-virial regions contain gas with broad, shallow density PDFs \citep[see also][]{burk18}.  

Figure \ref{fig:alphaAlphavir} shows how $\alpha$ in our fiducial model varies with virial parameter $\alpha_{\rm vir}$ measured in each of the plotted regions, overall exhibiting a stronger link than between $\alpha$ and $\langle\Sigma_{\rm mol}^{\rm cloud}\rangle$.  Also shown for reference is the line
\begin{equation}
\alpha=1.5+\frac{\mathcal{R_\alpha}}{\log\alpha_{\rm vir}}\label{eq:approximatealpha}
\end{equation}
chosen to asympotote to $\alpha=1.5$ at large $\alpha_{\rm vir}$.  
The constant scaling factor $\mathcal{R_\alpha}$ here is plotted in Fig. \ref{fig:alphaAlphavir} with the value $\mathcal{R_\alpha}=0.23\pm0.06$, chosen from the mean  and rms of $(\alpha-1.5)\log\alpha_{\rm vir}$ measured across the plotted data points.  

It should be kept in mind that, although the line in Fig. \ref{fig:alphaAlphavir} does a surprisingly good job of describing the measurements, it is purely empirical and applies only together with our nominal model.  It is a calibration of the relationship between $\alpha$ and $\alpha_{\rm vir}$ needed to match our nominal model for $\epsilon_{\rm ff}$ to the star formation in extragalactic clouds observed by PHANGS.  Still, it may offer a convenient, testable way to assign the $\alpha$ expected for a given $\alpha_{\rm vir}$, one based on observations of nearby galaxies that can be examined with observations and simulations.  

In Appendix \ref{sec:appendixefftrends} we show predictions of the sgMFF model that incorporate the PHANGS-calibrated relation between $\alpha$ and $\alpha_{\rm vir}$ plotted in Figure \ref{fig:alphaAlphavir}.  
We emphase that the trend between $\alpha$ and $\alpha_{\rm vir}$ % in Fig. \ref{fig:alphaAlphavir} 
(or between $\alpha$ and $R$ or $\langle\Sigma_{\rm mol}^{\rm cloud}\rangle$ in Figure \ref{fig:results1})
%$\langle\Sigma_{\rm mol}^{\rm cloud}\rangle$ 
 is a consequence of our choice to let variations in the power-law slope drive variations in $\epsilon_{\rm ff}$.  With this in mind, and to offer some validity to the predicted behavior of $\epsilon_{\rm ff}$ examined in Appendix \ref{sec:appendixefftrends}, below we discuss different physical pictures for how such a trend linking $\alpha$ to $\alpha_{\rm vir}$ might emerge.  
Here we first note that the selection biases are expected to have only a small role in shaping the trends recovered in  Figs. \ref{fig:results1} and \ref{fig:alphaAlphavir}.  We have confirmed that, in central environments, any clouds with higher $\alpha$ (less dense gas) than typical ($\alpha_{\rm ctr}\approx 1.6$) would be predicted with Eq. (\ref{eq:finaleff}) to form stars at rates that are still detectable.  Likewise, in the main disk, there appears to be a real absence of clouds with $\alpha$ below the typical $\alpha_{\rm disk}\approx 2$ and their associated higher rates of star formation.  

\subsubsection{Coordination of cloud evolutionary stage?}\label{sec:kvariation}
In the first possibility, the density structure reflects the physics of collapse and the variation in Figs.  \ref{fig:results1} and \ref{fig:alphaAlphavir} is the result of time evolution in the PL slope \citep[i.e.][]{kritsuk11,FK13,abreu,murray,caldwell18}. This would be remarkable as it would imply that there are  systematic variations in the average evolutionary stage of gas clouds throughout galaxies.    
Before discussing this possibility further, we note that the current model predictions explicitly neglect the possible time dependence of the PL slope and thus do not strictly account for time evolution.  However, as long as we focus on relatively late times in the evolution of the power-law, the full evolution can be neglected.  
Consider the scenario in which $\alpha$ asymptotes to $\alpha_f=1.5$. In this case, the further from $\alpha_f$ the gas is observed, the more a fit to Eq. (\ref{eq:finaleff}) will overstimate the true $\alpha$, or the degree of overestimation increases at earlier times.  To get a sense of the overestimation, we take a pure PL (Eq. \ref{eq:mffPL}) with a time dependent slope of the form $\alpha(t)=a_1 t^{-1}+\alpha_f$ and let $t_{\rm stop}\approx t_{\rm obs}\approx t_{\rm ff}(\rho_0)$.  Integration of the density PDF in the limit of large $s_{\rm crit}$ yields
\begin{equation}
\epsilon_{\rm ff}\approx\frac{\alpha_f-1}{\alpha-3/2}\left(\frac{\rho_{\rm crit}}{\rho_0}\right)^{(3/2-\alpha)}
\end{equation}
in the MFF scenario.  Thus, for $\alpha>3$, the true $\rm \epsilon_{\rm ff}$ is more than 4 times lower than the value predicted without time evolution in $\alpha$.  However, for any $\alpha$ currently in the range $1.5<\alpha<2.5$, the true $\rm \epsilon_{\rm ff}$ is within 1-3 times the time-independent prediction.  This translates into less than 10\% error in the $\alpha$ determined by fitting to $\epsilon_{\rm ff}$, from which we conclude that 
the fitted $\alpha$ should be a close approximation to the true current $\alpha$ for $1.5<\alpha<2.5$.   

With this in mind, the $\alpha$ values in Figs.  \ref{fig:results1} and \ref{fig:alphaAlphavir} determined by fitting the fiducial model to the measured $\rm \epsilon_{\rm ff}$ may indeed probe different stages in the evolution of clouds.  To match the systematic variation in $\alpha$ found in this work, the clouds at the innermost radii with low $\alpha$ and high $\alpha_{\rm vir}$ would need to be either more evolved than those at larger galactocentric radius or they would need to evolve systematically faster.  
In the former case, the low $\alpha$ in clouds at inner radii would imply that they have had more time to build up their high density interiors than their counterparts in the disk, i.e.,~because they survive longer.  This is typically opposite to the behavior of measured cloud lifetimes, which tend to be shorter where gas densities and velocity dispersions are higher (and thus the free-fall time and crossing time are also systematically shorter).  
Short free-fall times could alternatively imply faster evolution, so that central clouds are more often observed with high density tails.  This seems the reverse of the correlation implied in Fig. \ref{fig:results1} between $\alpha$ $\langle\Sigma_{\rm mol}^{\rm cloud}\rangle$ (tracking inverse free-fall time), but consistent with a weak link to $\sigma$, which is preferentially higher at small $R$ and responsible for the increasing departure from $\alpha_{\rm vir}=1$ there.  
This may suggest that shorter (turbulent) crossing times are capable of speeding up cloud contraction, helping to rapidly build shallow power-law tails.  Since cloud properties (including gas velocity dispersion) vary systematically with environment \citep{sun20,rosolowsky21}, this suggests that the galaxy may be playing a role in coordinating the cloud evolutionary state.  

\subsubsection{Impact of external pressure?}\label{sec:presssure}
In the second possibility, the galaxy's influence on cloud-scale gas motions has a more direct impact on internal structure through the virial state of the gas.  In particular, the increase in $\alpha_{\rm vir}$ (decrease in $\gamma$) that occurs with the decrease in $\alpha$ would signify an increasingly large external component to the gas pressure compared to self-gravity 
%\citep[e.g.][see also]{heyerketo}{sun20}.
(e.g.\ \citealt{heyer,field11}; see also \citealt{sun20}).
Numerical simulations suggest that clouds experiencing elevated external pressures contain shallower density profiles \citep{Anathpindika18}.  The decrease in $\alpha$ observed moving inwards to  dense galaxy centers where the external pressure is systematically elevated (see Fig.~\ref{fig:gammas}) is perhaps consistent with this expectation.    

A closely related scenario appeals directly to the gravitational forcing exerted by the galaxy to explain super-virial motions, rather than incorporating this influence as an external pressure, although the two are equivalent.  In this description, the internal structure of the gas is easily related to the virial level of the gas, as we sketch below.  

Consider motion in the rotating frame centered on a cloud orbiting around the galaxy with angular rate $\Omega$.  In this frame, the galaxy exerts a radial force $\kappa^2 r'$ a distance $r'$ from the center of mass (aligned with the galactocentric radius direction) and a vertical force $\nu^2 z$ at vertical distance $z$.  Here (and as given later in Appendix \ref{sec:appendixgamma}), $\kappa^2=\partial^2\Phi_{\rm gal,eff}/\partial r^2$ and $\nu^2=\partial^2\Phi_{\rm gal,eff}/\partial z^2$ where $\Phi_{\rm eff}$ is the effective potential of the galaxy in the rotating frame.  

In the cloud-centered rotating frame, the Euler equation of motion in the radial direction is 
\begin{equation}
\ddot{r'}=-\kappa^2r'-\frac{r'}{r}\frac{GM(r)}{r^2}-\sigma^2\frac{\partial \ln \rho_{\rm gas}(r)}{\partial r'}-\frac{\partial \sigma^2(r)}{\partial r},\label{eq:eqnofmotioncloudR}
\end{equation}
where the gas in the cloud has density $\rho_{\rm gas}$, pressure $P=\rho_{\rm gas} \sigma^2$ and
\begin{equation}
M(r)=4\pi \int_0^r \rho_{\rm gas}(r)r^2dr.
\end{equation}
Similarly in the $z$ direction\footnote{As discussed in $\S$ \ref{sec:M20bottleneck}, the galactic force terms in eqs. (\ref{eq:eqnofmotioncloudR}), (\ref{eq:eqnofmotioncloudZ}) and (\ref{eq:eqnofmotioncloudY}) lead to oscillatory motion within the cloud that sets up a velocity dispersion and can be revisioned as an effective pressure $P_{\rm gal}$. For motion in the $r'$ direction that obeys the  equation of motion $\ddot{r'}=-\kappa^2 r'$ (in the absence of gas self-gravity), for example, we set $\ddot{r'}=\partial P_{\rm gal}/\partial r'$ and rewrite the equation of motion as 
\begin{equation}
0=-\kappa^2r'-\frac{\partial P_{\rm gal}(r)}{\partial r'}.\label{eq:pressureeqnR}
\end{equation} 
Following \cite{meidt18} we can then define the effective velocity dispersion $\sigma_{\rm gal}^2=P_{\rm gal}/\rho$ for an assumed gas density distribution $\rho$ (see Appendix \ref{sec:appendixbk}).  Alternatively, we can determine the gas density profile associated with a given $\sigma_{\rm gal}$.  For a constant (isothermal) velocity dispersion, the equilibrium scenario represented by eq. (\ref{eq:pressureeqnR}) yields  
the density profile given in eq. (\ref{eq:edgeprofile}) treating the force balance (or equilibrium) in the vertical direction analogously to that in the radial direction.  }  
\begin{equation}
\ddot{z}=-\nu^2z-\frac{z}{r}\frac{GM(r)}{r^2}-\sigma^2\frac{\partial \ln \rho_{\rm gas}(r)}{\partial z}-\frac{\partial \sigma^2(r)}{\partial z}\label{eq:eqnofmotioncloudZ}
\end{equation}
and in the direction $y$ perpendicular to $r'$, 
\begin{equation}
\ddot{y}=-(2\Omega)^2r'-\frac{y}{r}\frac{GM(r)}{r^2}-\sigma^2\frac{\partial \ln \rho_{\rm gas}(r)}{\partial y}-\frac{\partial \sigma^2(r)}{\partial y}.\label{eq:eqnofmotioncloudY}
\end{equation}
These expressions can be used to determine the gas density profiles in the $r'$, $y$ and $z$ directions that achieve a balance of forces for an assumed velocity dispersion profile.  For comparison with the reference isothermal sphere, below we will consider isothermal gas.  (The last terms on the right hand sides of eqs. (\ref{eq:eqnofmotioncloudR}), (\ref{eq:eqnofmotioncloudZ}) and (\ref{eq:eqnofmotioncloudY}) can thus be neglected.)

Recall that, in the absence of the external forcing, the equation of motion is written
\begin{equation}
\ddot{r}=-\frac{GM(r)}{r^2}-\sigma^2\frac{\partial \ln \rho_{\rm gas}(r)}{\partial r}\label{eq:eqnofmotioncloudR}
\end{equation}
in terms of the conventional radial coordinate 
\begin{equation}
r^2=r'^2+y^2+z^2.
\end{equation}

In this case, the equilibrium solution is the cored isothermal density profile
\begin{equation}
\rho_{\rm gas}(r)=\frac{\rho_0}{1+\left(\frac{r}{a}\right)^2},\label{eq:sgprofile}
\end{equation}
which behaves like an isothermal profile $\propto r^{-2}$ outside a constant density core of size 
\begin{equation}
a=\left(\frac{\sigma^2}{4\pi G\rho_0}\right)^{1/2}.
\end{equation}

With the addition of the galactic force terms, the behavior of the equilibrium profile changes.  For example, taking the equation of motion in the $r'$ direction, we see that at $r'<a$, the constant density of the cored profile also satisfies equilibrium but now with
\begin{equation}
a=\left(\frac{\sigma^2}{\kappa^2+4\pi G\rho_0}\right)^{1/2}. \label{eq:reducedcore}
\end{equation}
The same relation follows in the $y$ and $z$ directions.  

The more prominent the galactic potential, the smaller the core at fixed $\sigma$.  As a result, more gas is shifted to high densities compared to the nominal equilibrium solution for a (non-singular) self-gravitating cloud.  
Outside the core, the equilibrium density profile can be expected to remain well-approximated by $r^{-2}$ as long as approximately $4\pi G\rho_{sg}/\kappa^2>1$, according to eqs. (\ref{eq:eqnofmotioncloudR}), (\ref{eq:eqnofmotioncloudZ}) and (\ref{eq:eqnofmotioncloudY}).  For larger distances, near the cloud edge and beyond, or roughly $r>a\sqrt{4\pi G\rho_{0}/\kappa^2}$, galactic forces dominate equilibrium.  As a result, the density in isothermal gas behaves like 
\begin{equation}
\rho_{edge}(r',z)=\rho_0e^{\frac{-r'^2}{2 r_0^2}}e^{\frac{-z^2}{2 z_0^2}}\label{eq:edgeprofile}
\end{equation}
for a given $\sigma$ where $r_0^2=\sigma^2/\kappa^2$ and $z_0^2=\sigma^2/\nu^2$.  Equation~(\ref{eq:edgeprofile}) is the equilibrium solution to Eqs. (\ref{eq:eqnofmotioncloudR}) and (\ref{eq:eqnofmotioncloudZ}) for isothermal gas in the limit of negligible self-gravity.   In this scenario, the density of the gas cloud asymptotes to zero and the mass approaches a finite value, relieving the description here from invoking an external pressure at the edge of a finite mass cloud and the associated Bonnor-Ebert profile.  

Compared to equivalent fully self-gravitating clouds, which are able to place more of their mass at large radii in equilibrium, clouds strongly influenced by galactic forces increasing outwards from the cloud center (proportional to  $r$)  necessarily contain more cloud mass at high density deep in their interiors.  The galaxy's reduction to cloud core in practice makes clouds overall more closely resemble $r^{-2}$ than when they are more fully self-gravitating, which would in turn make the power-law tail of the density PDF overall more shallow.  

The smaller core meanwhile contributes to the super-virial appearance of these clouds since the cloud mass and/or radius are changed with respect to their values in equivalent self-gravitating clouds with the same velocity dispersion.  Since the cloud mass $M_c=4\pi a^2\rho_0 R_c$ (integrating the cored equilibrium profile out to the cloud radius $R_c$ and looking in the limit $a/R_c\ll1$), then at fixed $R_c$ (fixed $M_c$) a smaller core radius translates to a smaller $M_c$ (larger $R_c$), both of which raise $\alpha_{\rm vir}$.  

From this perspective, the variation in internal structure and the change in $\alpha_{\rm vir}$ are both consequences of the variation in the galaxy's influence compared to gas self-gravity across disks. Note that we can envision the pressure needed to balance the weight of the gas in the galactic potential as an external pressure acting on self-gravitating clouds and reach the same conclusion.  Extragalactic studies that probe internal structure using proxies like HCN/CO do suggest that the dense gas content systematically increases with decreasing galactocentric radius \citep[e.g.][]{usero,gallagher18,Querejeta19,jimenez,neumann23,neumann25}, consistent with this picture.   

This close link between virial state and internal structure is only expected when the virial state is elevated specifically as a result of motion in the galactic potential, as assumed above. Other scenarios may also lead to an excess of cloud-scale kinetic energy ($\alpha_{\rm vir}>1$), e.g. when gas clouds and/or streams collide, gas accretion, the interaction or merging of host galaxies, star formation feedback.  In those scenarios, the additional forcing experienced by the gas may not elicit the same response as predicted in the presence of galactic forcing, with the consequence that the predicted density profile differs from either Eq. (\ref{eq:sgprofile}) or (\ref{eq:edgeprofile}).  It should thus be kept in mind that there are scenarios in which an elevated virial state may not necessarily imply that the gas contains a PL with $k$ approaching 2 ($\alpha\sim1.5$) and a relatively high amount of dense material (within which the critical density would also be relatively high; e.g. eq.[\ref{eq:scritKM05}]) as we find here (Figure \ref{fig:alphaAlphavir}).  Note, as well, that when the density PDF is closer to fully lognormal, it will contain even more intermediate-to-high density material than when a PL is also present (see $\S\S$ \ref{sec:powerlawchange} and \ref{sec:broadPL}).  This suggests that, outside the rotating gas disks of star-forming galaxies, the relation between virial state, internal structure, and dense gas content may be very different than we find here.  It is also likely evolving in time, at least partially reflecting the changing influence of the background galaxy.  

\subsection{Discussion: The impact of changes in $\alpha$ on MFF models and extragalactic star formation}\label{sec:discussionMFFvKM05}
The values of $\alpha$ implied by this work are capable of bringing MFF models into agreement with the observed rates of star formation in nearby PHANGS galaxies.  It is worth considering what those variations imply about the way galaxies form stars within the context of those models.  Recall that we introduce several modifications to MFF models that together reproduce the observed rate of star formation.  
The first of these is a broad power-law that accounts for self-gravity, compensates for changes in virial state, and in practice  helps overall reduce the output of the MFF models.  The second is the restriction to PDF renewal.  Those changes end up making MFF predictions behave, in many ways, closer to the original \citetalias{KM05} models.  
Consider, for one, the restriction to PDF renewal, which limits the MFF core formation process in time, making it behave similarly to the \citetalias{KM05} model where cores form only once in a cloud free-fall time, rather than continuously.  This is in addition to the flexibility added by inclusion of the power-law, which allows for changes in dense gas content that can, e.g., compensate for the penalty paid by gas in a super virial and weakly self-gravitating state.  
The result is that MFF star formation rates are kept from varying as widely as they would if the gas contained a uniform amount of the densest material, mimicking the variation of \citetalias{KM05} predictions and arguably improving the match to observations (see also Appendix \ref{sec:appendixefftrends}).  

While this might tend to suggest that the star formation process is better described by the \citetalias{KM05} model than  MFF models, context is important.  Whereas star formation efficiencies may exhibit relatively weak variations in star-forming galaxies \citep{leroy24}, where we suspect galactic regulation is active, in other systems the full flexibility of MFF models and the large dynamic ranges they can achieve may make them preferred (e.g.\ \citealt{salim15, kretschmerTeyssier, andalman24}; see section $\S$ \ref{sec:mainfeatures}).  
This includes in galaxies at earlier cosmic time ($z>2$), before a substantial stellar body has built up to take on its regulatory role, or in gas-rich starburst environments at low or high-z \citep[e.g][]{dz23}.

In this light, it is even more interesting that the changes in $\alpha$ found in this work imply that, in galaxies on the main sequence, the galaxy leaves little traceable imprint on $\epsilon_{\rm ff}$; at the same time as the galaxy alters the gas dynamical state, weakening the degree of self-gravitation, it coordinates the dense gas content to yield an overall roughly fixed star formation efficiency.   
Since the normal galactic disk environment is where the bulk of star formation since $z>2$ has occurred \citep{vdwel14}, this implies that the star formation process can generally be successfully modeled as taking place in decoupled, turbulent clouds at the basis of the \citetalias{KM05} model.  Again, incorporating the galaxy's influence in the context of a sgMFF model (as in this work), however, may be preferable when attempting to model the $\epsilon_{\rm ff}$ at much earlier cosmic time, or while simultaneously accounting for systematic variations in virial state and dense gas fraction that appear quite common in extragalactic observations.  

Altogether, the modifications proposed in this work ($\S$ 
 \ref{sec:proposedmodifications}) offer a natural way to renormalize the MFF models, by taking into account the self-gravity of the gas and the influence of factors in the environments of star-forming clouds that impact the PDF renewal at the heart of the MFF picture.  This includes feedback, magnetic fields, and galactic orbital motions (the latter being the specific focus of this work).  The resulting expression for $\epsilon_{\rm ff}$ (Eq. [\ref{eq:finaleff}]) is thus able to capture what is also found in numerical simulations \citep[e.g.][]{padoan12,federrath15,appel22,kretschmerTeyssier}, namely that star formation proceeds much more slowly than predicted by MFF models when the simulations capture ISM turbulence emerging over a range of scales and track it below the resolution \citep{semenov18, semenov24}.  
 
Numerical simulations also predict that the recovery of the adopted input efficiency may be sensitive to how efficiencies are (observationally) reconstructed \citep{grudic19,grisdale19,otero24}.  The time lag between when star formation occurs and when it is observed (given the timescale of the chosen tracer), as well as variations in cloud evolutionary state over kpc-scales, can impact how closely the efficiencies produced in the simulation match the adopted sub-grid efficiency \citep{otero24}.  In this light, the analytical model presented here may serve as a useful reference for distinguishing those effects from the impact that realistic galactic-scale turbulence has on on the efficiency of star formation.  In general, cosmological zoom-in simulations \citep[such as VINTERGATAN;][]{agertz21,renaud21a,segoviaotero22}, as well as idealized simulations of isolated galaxies 
\citep[see e.g.][]{gensior20,gensior21,gensior23,ejdetjarn22,renaud24} will continue to shed light on the environmental influences on the turbulent ISM, cloud-scale star formation, and their evolution over cosmic time.  

\section{Summary \& Conclusions}
Recent measurements of the cloud population-averaged (time-averaged) star formation efficiency per free-fall time $\epsilon_{\rm ff}$ throughout 67 nearby galaxies with PHANGS \citep[][see also Fig. \ref{fig:PHANGSsfe}]{leroy24} show variations with cloud properties that are in tension with most models of turbulence-regulated star formation.  
The implication is that the dependence of $\epsilon_{\rm ff}$ on turbulence Mach number and gas virial state are different (sometimes even reverse) than predicted by  theory (summarized in $\S$ \ref{sec:mainfeatures}).  In this paper we explore ways to improve the fit of the models to SF.  Our modifications ($\S$ \ref{sec:proposedmodifications}) build on a growing literature that relates variations in $\epsilon_{\rm ff}$ to the influence of a high density power-law tail that is missing from most models, but naturally develops as a consequence of gas self-gravity \citep{klessen, kritsuk11, murray,burk18,jaupart}.  
In these pictures, the $\epsilon_{\rm ff}$ is more sensitive to the slope of the power-law tail than to the turbulent Mach number.  

All of our modifications, which pivot on the inclusion of self-gravity, also capture the factors, especially those related to the galactic environment, that impact where gravity becomes the dominant factor influencing the evolution and renewal of gas density structure and its associated timescales.  We refer to the modified model as the self-gravitating MFF model or sgMFF. 

Like \cite{burk18}, in this model we adopt a hybrid density PDF,  but by lifting the requirement that it must be smooth (differentiable), we accommodate a power-law tail that is at once broad (extended) and as shallow as indicated by observations of local clouds \citep{Kainulainen14, schneider22}.  
For this more realistic broad power-law tail, the mass at high density is greatly reduced compared to either a pure lognormal PDF or the Burkhart smooth-PDF.  
As a result, it becomes much easier to match multi-free-fall models to the low efficiencies observed in local clouds and extragalactic targets without renormalization even when the star formation efficiency of dense cores inside the clouds reaches $\epsilon_{\rm core}=0.5$.   

As a physical factor that leads to a broad power-law tail, in this paper we consider the balance between gas self-gravity and the energy in motion in the background host galaxy potential ($\S$ \ref{sec:M20bottleneck}).  As described in $\S$ \ref{sec:decouplethresh}, we let the gas become self-gravitating at densities above where it kinematically decouples from galactic gravitational forces \citep[][estimated here in $\S$\ref{sec:potentials}]{meidt20}.  
For typical clouds in nearby galaxies, this transition takes place near the cloud edge, placing it notably in line with where resolved local clouds are shown to exhibit power-law behavior \citep[][see $\S$ \ref{sec:broadPL}]{Lombardi10,Lombardi15,Kainulainen14, alves17,schneider15,schneider22}.    

Another modification suggested by the view of gas self-gravity in the galactic bottleneck picture is related to the duration of PDF replenishment, which is assumed to be continuous in multi-free-fall models and lasting until at least a full free-fall time has elapsed.  
We propose that this duration can be shortened when gas in the outer cloud envelope is mostly coupled to the galaxy and unable to collapse, even if slowly ($\S$ \ref{sec:tstopmodel}).  That is, we let PDF renewal be dictated by self-gravity, and let the collapse timescale at the self-gravitating threshold set the duration of renewal (see \citetalias{HC11}). Star formation feedback and magnetic forces that can impact the dynamical state of the gas would also be factors that can impact the onset of self-gravitation and the replenishment of the density PDF \citep{girichidis14,appel22}.  Accounting for these factors, we limit the star formation process in the multi-free-fall scenario to some generic time $t_{\rm stop}$ that can be below the cloud free-fall time.  

With these choices and our adopted collapse threshold, the star formation efficiency predicted with our `self-gravitating' sgMFF model ($\S$ \ref{sec:obsTests}) is easily reduced to the low levels observed in the local universe which are also approximately matched by the LN SFF \citetalias{KM05} model.  By design, however, our chosen sgMFF model is also able match the higher efficiencies characteristic of starbursts and clumps at earlier cosmic time (\citealt{dz23}; and see \citealt{salim15} and references therein). In doing so, our model takes advantage of the added leverage of the power-law slope $\alpha$, which we find is the greatest factor impacting the precise value predicted for $\rm \epsilon_{\rm ff}$.  

 Matching PHANGS observations requires systematic variation in $\alpha$ across galaxies.  To obtain some insight into the processes that might be responsible for  variations in $\alpha$, we fit several models to the PHANGS data and solve for $\alpha$ ($\S$ \ref{sec:results}).  
Both our fiducial sgMFF model (broad PL plus limited MFF replenishment) and the smooth-PDF with SFF collapse proposed by \cite{burk18} can be fit with power-law slopes that are fully consistent with the range observed in local clouds.  The power-law behavior in the two models is otherwise quite different, however. 
 We quantify these differences using the fraction of gas above fixed density threshold (selected to be near the effective critical density for HCN) and the self-gravitating fraction as structural diagnostics.  We find that the broad PL PDFs entail considerably lower dense gas fractions than the smooth-PDFs where the lognormal component is more dominant.  This brings the former into closer agreement with extragalactic dense gas fractions probing comparable density contrasts \citep[e.g.][]{gallagher18,neumann23}.   

In either of the two best-fitting models, the power-law slope needed to match PHANGS measurements exhibits an increase with increasing $\langle\Sigma_{\rm mol}^{\rm cloud}\rangle$ and a decrease with increasing virial parameter.  Together with the sensitivity of cloud properties and virial state to galactic environment highlighted by recent work \citep{sun18,sun20a,liu21,sun22,rosolowsky21, meidt21,lu24}, this suggests that the host galaxy helps regulate cloud internal structure.  There are a few avenues for how this might occur, as discussed in $\S$ \ref{sec:discussion}.  For one, the total, deeper potential supports `super-virial' motions in the gas sometimes discussed in terms of an external pressure \citep[e.g.][see \ref{sec:presssure}]{meidt18,sun18,sun20}.  
According to the numerical study of \cite{Anathpindika18} this external pressure might then act to lower the power-law slope.  In support of this scenario, we examined the Euler equation of motion to infer how gas responds in the presence of galactic forces.  Since these forces increase with distance from the cloud center, more of the cloud mass is forced to high density in the cloud interior than characteristic of fully self-gravitating clouds.  
Clouds embedded in a strong background galactic potential appear closer to the isothermal $\alpha\approx 1.5$ expectation than their self-gravitating counterparts, which can contain a broader constant density core that steepens the density PDF power-law tail.      

Another possibility is that the galaxy coordinates the average cloud evolutionary state in any given region, preferentially giving clouds in galaxy centers the opportunity to build power-law tails either faster or for longer than in the main disk environment ($\S$ \ref{sec:kvariation}).  
Since cloud lifetimes tend to be shorter (rather than longer) in centers, we suggest that the enhanced non-thermal motions characteristic of galaxy centers, which translate into shorter turbulent crossing times, could signify that cloud evolution is sped up here, supporting rapid development of a power-law tail.   In this way, the time evolution predicted for the power-law slope \citep[e.g.][]{kritsuk11,FK13,girichidis14,abreu,murray,caldwell18} within clouds would have a recoverable influence on extragalactic star formation efficiencies.  

In either case, our results suggest that galaxy regulation may be key to modelling extragalactic star formation, even if the signs of this regulation may not always be apparent.  The two functions  we attribute to the galaxy -- setting the amount of dense material in gas that it simultaneously shifts to a super-virial, weakly self-gravitating state (restricting PDF renewal) -- have competing influences on $\epsilon_{\rm ff}$ ($\S$ \ref{sec:discussionMFFvKM05}).  The result, within the context of our nominal sgMFF models, is that the galaxy's influence on the star formation efficiency is imperceptible in practice.  
As a further consequence, clouds, in producing stars, behave nearly as envisioned in the \citetalias{KM05} model, i.e. as roughly virialized objects in which turbulence regulates star formation output in the course of a cloud free-fall time.  
Thus, despite outward signs that the gas is not fully self-gravitating and rather coupled to the galaxy (i.e. exhibiting deviations in virial state and internal structure), star formation in main sequence galaxies can be readily modeled as approximately virial, as envisioned by \citetalias{KM05}, arguably thanks to the galaxy's influence.  

For modelling star formation beyond the galaxy's regulatory influence, on the other hand, the full flexibility of the MFF class of models (with or without including a power-law component in the gas density PDF) may offer the preferable choice.  This includes starbursts with locally high gas fractions and at earlier cosmic time, before a substantial stellar body has built up.  
In both cases, with Eq. (\ref{eq:finaleff}) we predict considerably higher efficiencies than when the galaxy regulates the star formation process.  MFF models are also indispensable for predicting star formation efficiencies while simultaneously treating variations in $\alpha$ and virial state with galactic environment, as employed in this work.   

To test the role of the galaxy in the regulation of star formation speculated on here, it will be necessary to independently establish a connection between $\alpha$ and environment, i.e. through the study of extragalactic multi-line surveys that probe gas over a range of densities (covering from centers to disks).  Multi-scale numerical simulations of star-forming gas in the context of a fully evolving galaxy potential, with feedback and chemistry included, will also be key.  The ability to trace the full budget of star formation, e.g. with radio continuum observations, will help refine this view further, as will the recovery of star formation at early times and on small scales, e.g. with JWST, allowing for measurements of the efficiency at different stages to be uncovered.  Such data sets will shed continued light on the relationship between the star formation efficiency, gas structure and galactic environment.

\begin{acknowledgements}
We would like to thank the referee for a detailed review and insightful comments.  
This work was carried out as part of the PHANGS collaboration.

SCOG and RSK acknowledge financial support from the European Research Council via the ERC Synergy Grant ``ECOGAL'' (project ID 855130), from the German Excellence Strategy via the Heidelberg Cluster of Excellence (EXC 2181 - 390900948) ``STRUCTURES'', and from the German Ministry for Economic Affairs and Climate Action in project ``MAINN'' (funding ID 50OO2206). They are also grateful for computing resources provided by the Ministry of Science, Research and the Arts (MWK) of the State of Baden-W\"{u}rttemberg through bwHPC and the German Science Foundation (DFG) through grants INST 35/1134-1 FUGG and 35/1597-1 FUGG, and also for data storage at SDS@hd funded through grants INST 35/1314-1 FUGG and INST 35/1503-1 FUGG.

JG gratefully acknowledges funding via STFC grant ST/Y001133/1.

JDH gratefully acknowledges financial support from the Royal Society (University Research Fellowship; URF/R1/221620).

This paper makes use of the following ALMA data, which have been processed as part of the PHANGS--ALMA CO~(2-1) survey: \\
\noindent ADS/JAO.ALMA\#2012.1.00650.S, \linebreak % (N628/M74)
ADS/JAO.ALMA\#2013.1.00803.S, \linebreak % (N5128/CenA)
ADS/JAO.ALMA\#2013.1.01161.S, \linebreak % (N1365 + N5236/M83)
ADS/JAO.ALMA\#2015.1.00121.S, \linebreak % (N5236/M83)
ADS/JAO.ALMA\#2015.1.00782.S, \linebreak % (N1313 + N7793)
ADS/JAO.ALMA\#2015.1.00925.S, \linebreak % (pilot low mass)
ADS/JAO.ALMA\#2015.1.00956.S, \linebreak % (pilot high mass)
ADS/JAO.ALMA\#2016.1.00386.S, \linebreak % (N5236/M83)
ADS/JAO.ALMA\#2017.1.00392.S, \linebreak % (low mass follow-up)
ADS/JAO.ALMA\#2017.1.00766.S, \linebreak % (early-type)
ADS/JAO.ALMA\#2017.1.00886.L, \linebreak % (large program)
ADS/JAO.ALMA\#2018.1.01321.S, \linebreak % (N253, N300, Circinus)
ADS/JAO.ALMA\#2018.1.01651.S, \linebreak % (main sample follow-up)
ADS/JAO.ALMA\#2018.A.00062.S, \linebreak % (ACA-only nearby)
ADS/JAO.ALMA\#2019.1.01235.S, \linebreak % (local sample follow up)
ADS/JAO.ALMA\#2019.2.00129.S, \linebreak % (N1068)
ALMA is a partnership of ESO (representing its member states), NSF (USA), and NINS (Japan), together with NRC (Canada), NSC and ASIAA (Taiwan), and KASI (Republic of Korea), in cooperation with the Republic of Chile. The Joint ALMA Observatory is operated by ESO, AUI/NRAO, and NAOJ. The National Radio Astronomy Observatory is a facility of the National Science Foundation operated under cooperative agreement by Associated Universities, Inc.

\end{acknowledgements}

\appendix
\section{The background galactic potential: constraining where gas becomes self-gravitating}\label{sec:appendixgamma}
To empirically estimate the threshold density for self-gravitation $\rho_{\rm G}$ (where self-gravity dominates over the background potential) for each 'representative cloud' in a given measurement region, we need to calculate $\gamma$ in Eq. (\ref{eq:gamma1}).  This requires an estimate of the gas self-gravity (described in the main text) and the strength of the background galaxy potential on the outer cloud scale $R_{\rm c}$.  

As derived in \cite{meidt18} and \cite{meidt20}, the background (rotating) potential is written in terms of the effective pressure set up by the gas kinematic response to the associated gravitational forcing, which can be approximated with the angular velocity $\Omega$ and radial and vertical epicyclic frequencies $\kappa$ and $\nu$ defined as
\begin{eqnarray}
\kappa^2&=&\frac{\partial^2\Phi_{\rm eff}}{\partial r^2}\nonumber\\
&=&\left(4\Omega^2+r\frac{d\Omega^2}{dr^2}\right)\\
&=&2(1-\beta)\Omega \label{eq:kappa}
\end{eqnarray}
and
\begin{eqnarray}
\nu^2&=&\frac{\partial^2\Phi_{\rm eff}}{\partial z^2}\nonumber\\
&=&4\pi G \rho_{tot}-\kappa^2+2\Omega^2 ,\label{eq:eqnu}
\end{eqnarray}
where $\Phi_{\rm eff}$ is the galaxy effective potential in the rotating frame and $\beta$ is the logarithmic derivative of the rotation curve.   The expression in the second line of Eq.~(\ref{eq:eqnu}) follows from Poisson's equation, with $\rho_{\rm tot}$ the total density of the disk.  
 
For each galaxy we constrain these frequencies at all location of the PHANGS measurement grid of 1.5-kpc hexagonal regions ($\S$ \ref{sec:phangssfes}; see \citealt{sun22}) using the observed galaxy rotation curve and the stellar mass density, in addition to the observed gas properties.   Following \cite{sun23} we adopt Legendre polynomial fits (Nofech et al., in prep.) to each of the \cite{lang20} rotation curves measured for each galaxy from the observed CO kinematics in a series of 150pc-wide radial bins. 

These polynomial fits allow us to avoid a noisy numerical derivative of $\Omega^2$ (see Eq. Eq.~(\ref{eq:kappa}).  For each each hexagonal region we assign the value for $\beta$ obtained from each polynomial curve at that location, as tabulated by \cite{sun23}.  Each region is assigned a single value $\kappa$ and $\Omega$ (and $\nu$) neglecting variation in these quantities across the hex area.  

Our estimate of the vertical epicyclic frequency is derived using Eq.~(\ref{eq:eqnu}) in terms of $\Omega$, $\kappa$ and the mid-plane background density, which we assume is dominated entirely by the stellar component.  As our focus is on the inner, molecule-rich star-forming parts of the disk, we ignore dark matter and the contribution from atomic gas and thus assume that $\rho_{\rm tot}=\rho_{\rm stars}+\rho_{\rm mol}$ where $\rho_{\rm stars}$ is the background stellar volume density.  

To obtain an estimate for $\rho_{stars}$ at all locations, we follow \cite{sun20} and adopt the PHANGS 2D stellar mass surface density maps and assume a constant scale height $z_0$, selected using the empirical scaling relation measured by X between scale height and stellar mass.  The assumption of a flared stellar disk can have a modest impact on our assessment of the relative strength of self-gravity 
\citep[see also][]{sun20}.  However testing suggests that this does not majorly significant impact the trends reported in section \ref{sec:results}.  

\section{derivation of the density factor \lowercase{$b_k$}}\label{sec:appendixbk}
\cite{meidt20} present an estimate for the component of the velocity dispersion in clouds due to orbital motion in the galactic potential.  Considering the spread in orbital motion  within extent $x_0$ in a cloud, the velocity dispersion throughout an entire cloud can be obtained by integrating these motions across the full extent of the cloud, 
taking into account its internal density structure. 

For simplicity, we work in a regime in which $\kappa\approx\sqrt{2}\Omega$, which is valid in disk locations with an approximately flat rotation curve, so that clouds in dynamical equilibrium with the background gravitational potential are symmetric in the plane \citep[see][]{meidt18}.  As in the main text, we allow the cloud to be flattened (triaxial), with vertical to radial axis ratio $q$.  In this case, a model for the gas density that is not spherically distributed seems appropriate, and we adopt a power-law density profile
\begin{equation}
\rho=\rho_0\left(r^2-\frac{z^2}{q^2}\right)^{-\frac{k}{2}}.\label{eq:triaixialrho}
\end{equation}

The gas parcels that make up a cloud embedded in an external potential were described in \cite{meidt18} as exhibiting a spread in velocity that depends on distance from the cloud center, i.e. 
\begin{equation}
\left<v_x^2\right>\approx\left<v_y^2\right>\approx\kappa^2r^2
\end{equation}
and
\begin{equation}
\left<v_z^2\right>\approx\nu^2 z^2
\end{equation}
in the cloud frame, now in terms of a parcel's average squared position $r^2=x_0^2/2$ in the plane and $z^2=z_0^2/2$ in the vertical direction.  This constitutes a velocity dispersion within the cloud that can be calculated by 
performing the following integrals over the cloud volume 
\begin{equation}
\sigma_{plane}^2=\frac{\int_0^{qRc}\int_0^{\sqrt{R_{\rm c}^2-z^2/q^2}}\rho (\kappa^2 r^2) r dr dz}{\int_0^{qRc}\int_0^{\sqrt{R_{\rm c}^2-z^2/q^2}}\rho r dr dz }
\end{equation}
and 
\begin{equation}
\sigma_z^2=\frac{\int_0^{qRc}\int_0^{\sqrt{R_{\rm c}^2-z^2/q^2}}\rho \nu^2 z^2 r dr dz}{\int_0^{qRc}\int_0^{\sqrt{R_{\rm c}^2-z^2/q^2}}\rho r dr dz }
\end{equation}
with $\rho$ given by Eq.~(\ref{eq:triaixialrho}).  
These yield 
\begin{eqnarray}
\sigma_{plane}^2=b_{k}2\kappa^2 R_{c}^2\\
\sigma_z^2=b_{k}\nu^2 z_{c}^2,%/q^2
\end{eqnarray}
where 
\begin{equation}
b_{k}=\left(\frac{3-k}{5-k}\right).
\end{equation}
Thus the total 1D velocity dispersion due to galactic motion in the gas is 
\begin{equation}
\sigma_{gal,1D,r}^2=\frac{b_{k}}{3}\left(2\kappa^2 R_{c}^2+\nu^2z_c^2\right)
\end{equation}
in the rotating frame when $\kappa^2\approx 2\Omega^2$.  

In the non-rotating (inertial) frame, the 1D velocity dispersion is slightly different, but the density factor $b_k$ is the same, 
\begin{equation}
\sigma_{gal,1D,i}^2\approx\frac{b_{k}}{3}\left(\frac{1}{2}\kappa^2 R_{c}^2+\nu^2z_c^2\right)
\end{equation}
when $\kappa^2\approx 2\Omega^2$.  

%\section{Extragalactic dense gas fractions}
\section{Predictions of the PHANGS-calibrated sgMFF model }
\label{sec:appendixefftrends}
\begin{figure*}[t]
%\begin{flushleft}
\vspace*{-.5in}
\begin{center}
\begin{tabular}{c}
\includegraphics[width=0.85\linewidth]{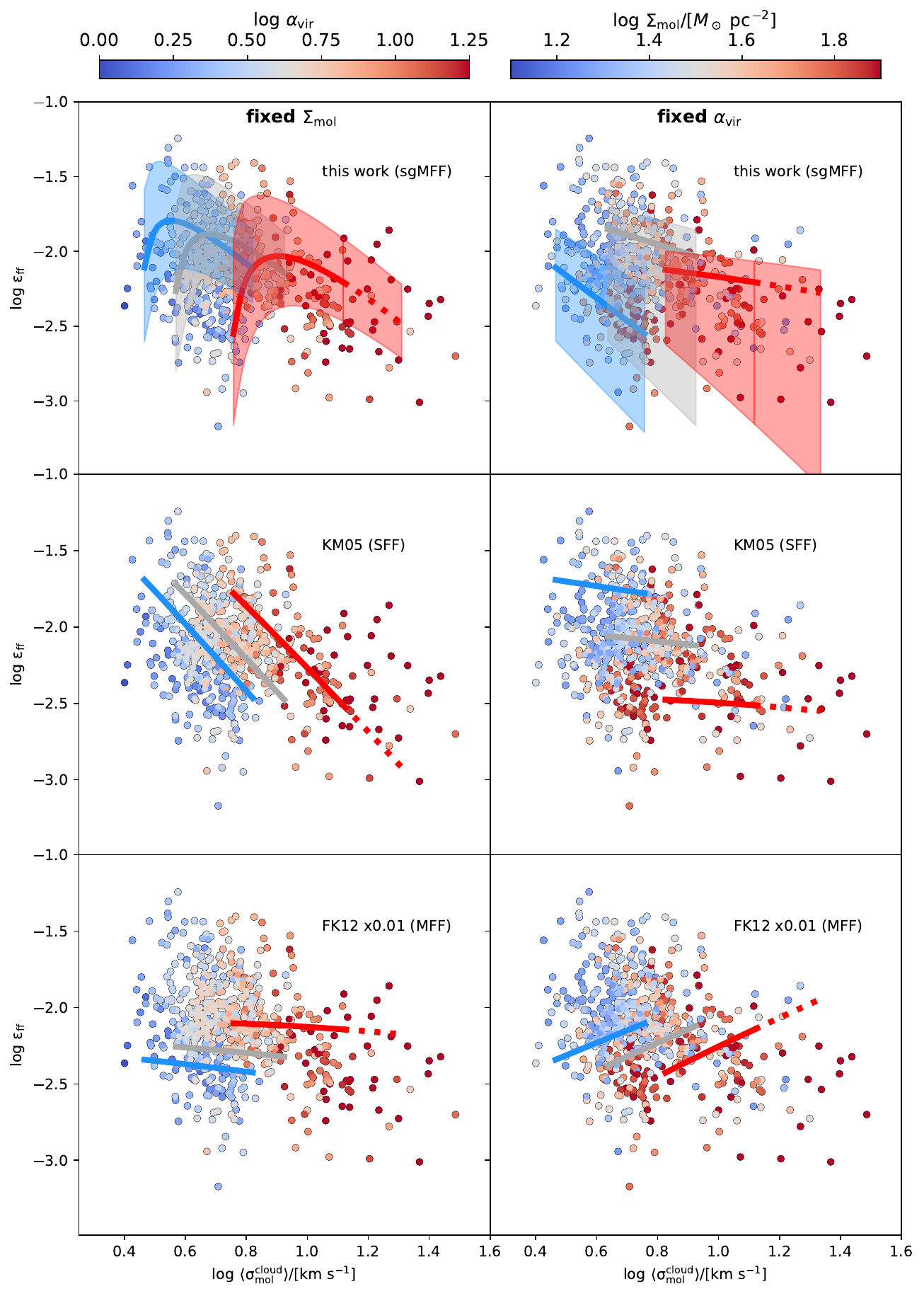}%phangs_models.jpg}%modBplot150nh.jpg}%pdf}
%\vspace*{-.5in}
\end{tabular}
\end{center}
%\vspace*{-.15in}
\caption{Comparison between 3 star formation models (top, middle and bottom) and PHANGS measurements of the population-average $\rm \epsilon_{\rm ff}^{\rm obs}$ in 1.5-kpc wide hexagonal apertures (see Figure 1) plotted against the average cloud-scale velocity dispersion $\langle\sigma_{\rm mol}^{\rm cloud}\rangle$ in each aperture (see text). In the left (right) panels, measurements are color-coded by $\langle\Sigma_{\rm mol}^{\rm cloud}\rangle$ ($\langle\alpha_{\rm vir}\rangle$) and shown alongside lines illustrating fixed-$\Sigma$ (fixed-$\alpha_{\rm vir}$) predictions from turbulence-regulated SF models at three representative values of $\Sigma$ ($\alpha_{\rm vir}$) (see text).  
As in Figure 1, all models adopt $\epsilon_{\rm core} =0.5$, $b=0.87$, and gas sound speed $c_s = 0.3 \, {\rm km \, s^{-1}}$ (see main text for definitions of these quantities) and, for comparison purposes, the same definition for the critical density, given by Eq. (\ref{eq:scritKM05}).   
 }
\label{fig:PHANGSsfeMODEL}
%\end{flushleft}
\end{figure*}
In this section we illustrate the trends in $\epsilon_{\rm ff}$ vs. cloud scale properties predicted by Eq.~(\ref{eq:finaleff}) when using the calibration in Eq. (\ref{eq:approximatealpha}) between $\alpha$ and $\alpha_{\rm vir}$ suggested in $\S$ \ref{sec:discussion} from the match between the sgMFF model and the PHANGS measurements.  Substituting instead the $\alpha$ values fitted region by region (plotted in Figure 6) back into the model would (by design) yield predictions for $\epsilon_{\rm ff}$ that match the observations as closely as possible.\footnote{The agreement is not perfect as a result of the grouping strategy we use during fitting, which assigns the $\alpha$ determined for any given group of regions to each of the regions in that group.}  
The calibration in Eq. (\ref{eq:approximatealpha}) makes it possible to assess how well the calibrated model reproduces the data in comparison to other conventional turbulence-regulated star formation theories.  

Figure \ref{fig:PHANGSsfeMODEL} here is similar to Figure 1 and examines predictions for $\epsilon_{\rm ff}$ from three different models as a function of $\sigma$ both at fixed $\alpha_{\rm vir}$ (left panels) and at fixed $\Sigma$ (right panels).  The three models are: the calibrated sgMFF model proposed in this work (top), the KM05 SFF model (middle), and the FK12 MFF model (bottom) scaled by a factor of 1/100.  In each panel, three trend lines are drawn at three different representative values, chosen from the 10th, 50th and 90th percentiles of the observed distributions of either $\langle\Sigma_{\rm mol}^{\rm cloud}\rangle$ or $\langle\alpha_{\rm vir}\rangle$ across the full set of measurements (see Figs.~\ref{fig:results1} and \ref{fig:alphaAlphavir}), to illustrate the average and approximate range of the observations.  Each trend line at fixed $\alpha_{\rm vir}$ ($\Sigma$) is then extended in $\sigma$ only as far as allowed by the low and high representative values for $\Sigma$ ($\alpha_{\rm vir}$).  Line lengths could be chosen to depict the full range of the data (rather than the range from the 10th to the 90th percentile).  For illustration, we include a segment in each plot that extends the predictions for the highest $\alpha_{\rm vir}$ (highest $\Sigma$) trend lines out to the value for $\sigma$ corresponding of the 99th percentile of the distribution of $\langle\Sigma_{\rm mol}^{\rm cloud}\rangle$ ($\langle\alpha_{\rm vir}\rangle$).  

As noted previously, the relation between $\epsilon_{\rm ff}$ and $\sigma$ (rather than $\Sigma$ or $\alpha_{\rm vir}$) is our preferred point of reference, as the measurements $\epsilon_{\rm ff}^{\rm obs}$ and $\langle\sigma_{\rm mol}^{\rm cloud}\rangle$ that we will compare against are most independent of each other.  As emphasized by \citealt{leroy24}, the correlation introduced between measurements of $\epsilon_{\rm ff}^{\rm obs}$ and $\langle\Sigma_{\rm mol}^{\rm cloud}\rangle$ (or $\langle\alpha_{\rm vir}\rangle$ involving $\langle\Sigma_{\rm mol}^{\rm cloud}\rangle$), which are not independent, makes interpreting the relation between these variables less straightforward.  

In this illustration, we set the basic model parameters shared by the models ($\epsilon_{\rm core} =0.5$, $b=0.87$, and gas sound speed $c_s = 0.3 \, {\rm km \, s^{-1}}$) to the values used in Figure 1, and select the same critical density $s_{\rm crit}$ given by Eq. \ref{eq:scritKM05} for all models.  Note, though, that as in Figure 1, the KM05 and FK12 predictions adopt $\phi_t=1.9$, whereas the sgMFF model adopts $\phi_t=1$.  As noted previously, variations in these basic parameters are not expected to greatly extend the range of the predictions.  The addition of the PL in the sgMFF model, on the other hand, gives predictions from this model an added degree of freedom compared to either KM05 or FK12.  We opt to show 
each sgMFF trend line surrounded by a band that represents the rms spread in $\alpha$ around the average trend fitted to the measurements in Figure \ref{fig:alphaAlphavir} (reported in the text).  

The predictions in Figure \ref{fig:PHANGSsfeMODEL} exhibit a number of interesting features.  First, although the dynamic range of the FK12 model is overall larger (see Figure 1), over the observed range of cloud properties in the PHANGS-ALMA sample under consideration, this model exhibits the least variation in $\epsilon_{\rm ff}$.  The three trend lines do not encompass the observed range in efficiency as well as either the KM05 SFF or the sgMFF model although, in both left and right panels, each trend line sits along the horizontal average near enough to similar-valued measurements.  

In comparison, the KM05 model does a better job of spanning the range in $\epsilon_{\rm ff}^{\rm obs}$ given the observed gas properties.  The trend lines sit near the positions of similar-valued measurements in both the horizontal and vertical axes.   However, not all of the features of the measurements are reproduced by the predictions.  In the middle right panel, for example, regions with low $\alpha_{\rm vir}$ exhibit a wider range in $\epsilon_{\rm ff}^{\rm obs}$ than predicted and some of the high $\alpha_{\rm vir}$ regions appear to form stars more efficiently than in the model.  In the middle left panel, the KM05 model also falls short of predicting the lowest (highest) efficiencies at low (high) $\Sigma$.  

As shown in the top row, the sgMFF model performs at least as well as the KM05, starting with a foundation that is more similar to the FK12 model.  As discussed in $\S$ \ref{sec:discussionMFFvKM05}, elements of both of those models are reproduced by sgMFF, although the addition of the PL yields even greater flexibility to match the observations.   Indeed, many of the shortcomings of the KM05 (or FK12) model can be overcome since, by design, the sgMFF model in this instance has been calibrated to reproduce many of the features at fixed $\Sigma$ and $\alpha_{\rm vir}$ that were not well reproduced by KM05 or FK12.  
In the left panel, for example, fixed $\Sigma$ trends capture a wider range of $\epsilon_{\rm ff}$ and now also show a slightly larger decrease in efficiency with increasing $\Sigma$ that was only modest in KM05 and absent in FK12.  In addition, predictions from sgMFF at low $\alpha_{\rm vir}$ in the top right panel are shifted down compared to the KM05 prediction (middle right panel), correcting for the overestimation in $\epsilon_{\rm ff}$ in the KM05 model there.  

It should be kept in mind that the model predictions shown here are tailored to apply under the conditions in the gas observed by PHANGS-ALMA.   Predictions would vary when applied to high-z objects, for example, which we have thus omitted from the present comparison. For such cases, the relation between virial state and internal structure is likely different than prescribed by the relation between $\alpha$ vs. $\alpha_{\rm vir}$ calibrated using PHANGs observations of the gas in extragalactic star forming disks.  
To match the high-z points appearing in Figure 1 with sgMFF model, the power-law index $\alpha$ would need to more closely approach $1.5$ (see Figure \ref{fig:newmodelsfe}), placing the gas in those objects very near a state of pure free-fall collapse.  Alternatively, the self-gravitating component would need to appear only at densities above the critical density, so that the core formation efficiency is strongly tied to the LN part of the PDF and boosted as in the original HC11 and FK12 models.

\end{document}